# Temperature Dependent Separation of Metallic and Semiconducting Carbon

*Iskandar Yahya[1]‡, Francesco Bonaccorso[2,3], Steven K. Clowes[1], Andrea C. Ferrari[2], and S. R. P. Silva[1]\**

[1] Advanced Technology Institute, University of Surrey, Guildford GU2 7XH, UK.

[2] Cambridge Graphene Centre, Cambridge University, Cambridge CB3 0FA, UK.

[3] Materials Characterization Facility, Istituto Italiano di Tecnologia, via Morego 30, 16163 Genova, Italy.

‡ Currently at the Institute of Microengineering and Nanoelectronics, Level 4, Research Complex, Universiti Kebangsaan Malaysia, 43600, Bangi, Malaysia.

\* Corresponding author: s.silva@surrey.ac.uk.

## ABSTRACT

Post-synthesis separation between metallic (m-SWNTs) and semiconducting (s-SWNTs) singlewall carbon nanotubes (SWNTs) is remaining a challenging process for the reliable fabrication of high performance electronic devices. Gel agarose chromatography is emerging as an efficient and large scale separation technique. However, the full (100%) separation of m-SWNTs and s- SWNTs has not been reached yet, mainly due to the lack of understanding of the separation mechanism. Here we study the temperature effect on the separation via gel agarose chromatography by varying the separation process temperatures between 6°C and 50°C, for four different SWNT sources. Exploiting the gel agarose micro-beads filtration technique we achieved up to 70% m-SWNTs and more than 90% s-SWNTs, independent of the source material. The process is temperature dependent, with improved yields up to 95% for s-SWNT (HiPco) at 6 °C. Temperature affects the sodium dodecyl sulfate SDS surfactant-micelle distribution along the SWNT sidewalls, thus helping to act as an aid in the sorting of SWNTs by electronic type. The sorted SWNTs are then used to fabricate transistors with very low OFF-currents ($\sim 10^{-13}$ A), high ON/OFF current ratio ($>10^6$) and improved charge carriers mobility ($\mu \sim 40 cm^2 V^{-1} s^{-1}$).

## INTRODUCTION

Carbon nanotubes (CNTs) have attracted much attention in view of their potential applications, which can exploit their transport properties [1] ranging from large area networks [2], to sensors [3] and individual SWNT devices [4], [5], [6], [7], just to cite a few.

However, applications often require SWNTs of specific electronic type: either metallic (m-SWNTs) or semiconducting (s-SWNTs). To date, the vast majority of synthesized SWNTs are composed of a mixture of the two electronic species. Growth of pure (100%) m-SWNTs or s-SWNTs is still being pursued. Recent works [8], [9], [10], [11], [12], [13], [14], [15], [16] reported SWNT enrichments based on electronic type [8], [9], [10], [11], [12] and specific chiralities [15], [16]. However, achieving on-demand control of SWNTs type is still far from being solved. The rise of grapheme [17], [18], [19] and other two-dimensional (2d) crystals [19], [20], [21], [22], has added further pressure to realize the promise of SWNTs in electronic devices, since 2d crystals do not require a sorting process to define their electronic nature (although some 2d crystals, *e.g.* $MoS_2$, show indirect band gap in bulk and direct in monolayer configuration [20], [21]), a step otherwise necessary in the case of SWNTs [23], [24], [25], [26], [27]. On the other hand, selectivity of SWNTs is not always required and, *e.g.*, the heterogeneity of un-sorted SWNTs can be exploited in ultrafast and mode-locked lasers [28], [29], allowing wideband tuneability, due to the presence of a variety of tube diameters and chiralities in a given sample [30].

Post-synthesis separation of m-SWNTs and s-SWNTs [23], [24], [25], [26], [27], which we will refer to as MS-separation, is an effective alternative and/or complementary tool to selective growth [8], [9], [10], [11]. A variety of different methods such as: ac-dielectrophoresis [23], DNA wrapping [24], selective breakdown of m-SWNTs [25], density gradient ultracentrifugation (DGU) [26], [31], [32], amine extraction [33] and gel agarose electrophoresis [27] have been exploited for such goal. These separation strategies have been reviewed in detail in Refs. [34] and [35]

Although DGU is by far the most complete separation strategy, and to date has permitted the separation by length [36], diameter [37], metallic vs. semiconducting [26], chirality [31], and handedness [38], it is time and resource consuming, and has low throughput at laboratory scale [37], [39]. On one hand it is possible to scale-up the DGU process and improve yield using large capacity centrifuges, but the costs involved are considerably high. Routinely, DGU separations require up to 20 hours [26], [37], [32] centrifugation time, which reflects on energy consumption. Other separation methods based on chemical approaches [40] involves irreversible covalent functionalization [41] with chemicals or biological impurities that not only affect the intrinsic SWNT properties [34], [42], [43], but may also have health and environmental impact [44], [45].

An alternative to DGU is separation via agarose gel chromatography. Ref. [46] reported that agarose gel beads are very promising for MS-separation, due to simplicity, low cost, short process time and scalability [46]. The processing time for gel agarose beads separation is ~20 minutes [34] compared to ~10-20 hours [26], [37] needed for the DGU. The agarose beads filtration requires only a single step and, as in the case of DGU [26], [31], the SWNTs dispersion is in aqueous surfactant solutions. Moreover, the separation purity level (*i.e.* the percentage of m- and/or s-SWNTs) is now comparable (*e.g.* ~ 90% for *(6,5)*, *(7,6)*, *(8,6)*) [47] with DGU (>95% both for m-SWCNTs [31] and s-SWNTs [26], [27]). It is possible to achieve purities up to~99% by multiple agarose gel filtration iterations [48], but the process is time and resource heavy. However, although the mechanism of separation by hydrogels (*e.g.,* agarose and sephacryl) was demonstrated to be an entropy-driven process [49], it is not fully understood yet. Furthermore, there are still other unexplored process parameters, such as temperature, which is crucial in order to optimize separation purity of this process.

Here, we report MS-separation of SWNTs from four different sources using gel agarose beads filtration. We investigate the effect of temperature on separation purity, showing that this affects the purity of sorted SWNTs. To the best of our knowledge, there is no prior studies focusing on the temperature dependence of the separation at an extended temperature range, i.e. 6 °C to 50 °C. Moreover, the separation is shown to be influenced by the surfactant micelles formation around the SWNTs' surface, which in turns is affected by temperature. From our study, we demonstrate that there is a trade-off between the purity percentage of the enriched sSWCNTs and m-SWCNTs fractions for separation carried out at varying temperature. Here, the purity percentage of s-SWCNTs is the highest at low temperature (~6 °C) and gradually drops as the temperature is increased (~50 °C), whereas for m-SWCNTs, the reverse behavior was observed. To demonstrate a direct application as an electronic device, the enriched s-SWNTs are then used to fabricate CNT-based field effect transistors (CNT-FETs) with superior performance in terms of ON/OFF current ratio (*e.g.* >$10^6$) with respect to devices based on un-sorted SWNTs (<2).

## EXPERIMENTAL METHODS

### Nanotube Sources

Four types of SWNTs are used. The first is prepared by enhanced direct injection pyrolytic synthesis [50], [51] (DIPS-CNT). These SWNTs have diameters in the range of 0.83nm, and are obtained from *Nikkiso Corp. Japan*. The second is arc-discharge SWNT (AD-CNT) from *Nanocarblab*, with diameters in the 1.2-1.4nm range [52]. The third is CoMoCAT SWNTs, with diameter distribution between 0.7-0.9nm [53], [54]. The fourth is high pressure carbon monoxide (HiPco) SWNTs [55], with 0.8-1.2nm diameter [55]. All SWNTs are from raw material, except for DIPS-CNT, which is purified by the supplier and contains >90% SWNTs.

### Dispersion

SWNT dispersions are prepared by adding 1mg/10ml weight to volume ratio of each SWNT source to an aqueous solution of 2% sodium dodecyl sulfate (SDS, *Sigma-Aldrich*) in deionised water (DIW). De-bundling is obtained via ultrasonication using a *Cole Parmer Ultrasonic Processor* for 4 hours (40% of 750 kW at 20 kHz). The obtained dispersions are ultracentrifuged using the *Beckman L8-70M Ultracentrifuge* at 200,000 x *g* for 20 minutes at 18°C to remove SWNT bundles and other impurities, such as amorphous carbon, catalyst residual etc [31]. The supernatant of the four obtained dispersions after the ultracentrifugation step is collected using pipettes and used for the gel filtration separation process.

*Separation Process*

Disposable 2.5 ml syringes, used as continuous separation columns, are plugged with cotton, and filled with 1.5 ml agarose micro-beads suspended in ethanol (Sepharose 2B, 2% Agarose, 60–200μm, *Sigma-Aldrich*). The syringe column is then washed and equilibrated with the surfactant aqueous solution by pouring 5 ml of 1% SDS in DIW into it and letting the solution pass through the agarose gel beads to remove the ethanol. ~0.5ml SWNT dispersion is then poured into the column. Successively, 1.5 ml 1% SDS solution is added. This causes a displacement of the SWNT dispersion along the column. A portion of the SWNTs are trapped at the top layer of the agarose beads, creating the top band, and another portion moves down the column, forming the bottom band, as for Figure 1. The bottom band is eluted from the column when an additional 1 ml of 1% SDS is added and then collected for spectroscopic characterization. Depending on the SWNT source, the top and bottom bands formed in the column have different colors. A further 1 ml 1% SDS solution is then added until the entire bottom band is eluted. To elute the top band, 3.5 ml of 1% sodium deoxycholate (SDC, *SigmaAldrich*) solution in DIW is poured into the column. In literature, other surfactants, such as sodium cholate (SC), Triton x-100, Tween 20 and sodium dedocylbenzene sulfonate (SDBS) have been tested for the elution of the top band [46]. Here, we use SDC due to its suitability for dispersing individual SWNTs [31], [46].

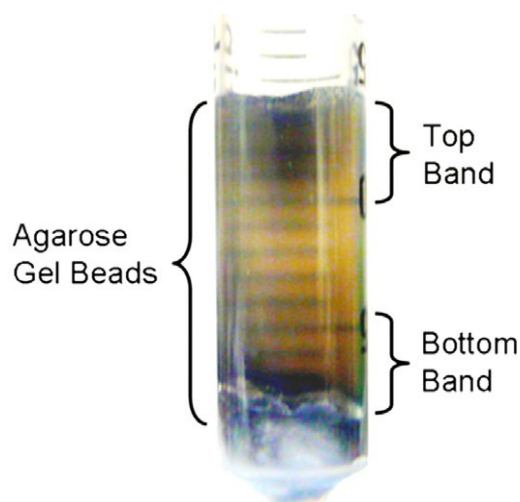

**Figure 1:** Syringe column filled with agarose gel beads. The dark colour shades are the SWNTs under separation in the column forming the top and bottom bands. The example shown here is the separation of a HiPco sample.

To study the effects of temperature, the process described above is repeated at different temperatures: 6, 10, 18, 24 (set as Room Temperature, RT), 30, 40 and 50 °C for the 4 sets of dispersions. The temperature of both SWNT dispersions and surfactant solutions (1% SDS and 1% SDC) is adjusted by heating/cooling them in a water bath, prior to use in the separation process. The temperature of the agarose beads in the syringe column is adjusted via pre-washing with the temperature-treated surfactant solutions. The last two steps were carried out to control the separation process temperature and minimize temperature variation.

*Characterization*

The characterization is carried out by optical absorption, photoluminescence excitation (PLE), and Raman spectroscopy.

*Optical absorption spectroscopy*

Optical absorption spectra (OAS) reveal various properties of SWNT dispersions, such as transition energies [56], [57], [58], bundling [57], [58], [59], and concentration [60]. The determination of the relative concentration of chiralities of sorted sample is carried out taking the derivative of absorption, as done in Refs. [26] and [31] to determine the peak absorption. For quantitative analysis, if we assume that the absorption spectrum obeys the Beer-Lambert law [61], $A_\lambda=\alpha_\lambda lc$, where $A_\lambda$ is the absorbance at wavelength $\lambda$ [nm], $\alpha_\lambda$ [L mol$^{-1}$ cm$^{-1}$] the absorption coefficient at the same wavelength, $l$ [cm] the length of optical path and $c$ [mol L$^{-1}$] is the concentration of material, there will be a similar relationship between concentration and absorbance for the first-order derivative: $\frac{dA_\lambda}{d\lambda} = \frac{da_\lambda}{d\lambda} lc$

The OAS are acquired in a Varian Cary 5000. Measurements are carried out in the range 400-1300nm, limited by the strong absorption of water. This range is sufficient to cover the first and second excitonic transitions of s-SWNTs [56], [62], [63] and the first transition of m-SWNTs, for CoMoCAT and HiPco [56], [62], [63]. For DIPS-CNT and AD-CNT, this covers the second and third transitions of s-SWNTs and first of m-SWNTs [64], [65]. Absorption from solvent and surfactants is subtracted, by measuring solutions with only solvent and surfactant.

The assignment of the optical transitions is based on the empirical Kataura plot of Ref. [64]. This gives values of optical transition frequencies versus chirality for SWNTs in aqueous surfactant dispersions, and is more appropriate than Kataura plots theoretically derived from tight binding and other models [64], [66]. We use the empirical Kataura plot of Ref. [64] also for chirality assignment in PLE and Raman spectroscopy.

*Photoluminescence excitation spectroscopy*

PLE is one of the most used techniques to monitor SWNT dispersions [66], [67], [68]. The ($eh_{ii}$,$eh_{11}$) resonances ($i$=1, 2, ... *etc*) from different SWNTs appear as sharp features ($\lambda_{ex}$, $\lambda_{em}$), where $\lambda_{ex}$ and $\lambda_{em}$ are the excitation and emission wavelengths, respectively. Other peaks can be observed, either due to excitonic-phonon sidebands [69], exciton energy transfer (EET) [57], [70] or bright phonon sidebands (BS) of dark K-momentum excitons [71].

In general, the PL intensity is proportional to the concentration of a species, its absorption cross-section [72] at the excitation wavelength [73], and its fluorescence quantum yield [74]. Thus, it does not directly reveal the relative abundance of SWNTs [73]. However, the relative PL intensity of different chiralities can be used to compare the separation process at different temperature. The PLE maps of the dispersions are recorded using a Horiba Jobin-Yvon excitation-emission spectrofluorometer (Fluorolog3) with 10 nm slits for the double grating excitation monochromator and 14 nm for the single grating emission. The scan step is 5 nm for excitation, with a range from 440 to 850 nm for CoMoCAT and HiPco SWNTs, and from 350 to 700 nm for DIPS-CNT samples, respectively. The emission is collected by a liquid-nitrogencooled InGaAs detector using a right angle scattering mode, in the ranges: 900-1200 nm for CoMoCAT, 850-1450 nm for HiPco and 1250-1550 nm for DIPS-CNT. PLE measurements are not carried out for AD-CNT due to the large diameter distribution of the sample that prevents the emission to be detected by the current configuration. Note that ~1600 nm is our upper detection limit. This restricts the detection of PL from SWNTs with diameter larger than ~1.3 nm [64]. The raw PL data are corrected by the excitation profile.

*Raman spectroscopy*

Raman spectroscopy can be used to probe SWNTs within dispersions. In the low frequency region, the Radial Breathing Modes (RBMs) are observed [75]. Their position, Pos(*RBM*), is inversely related to SWNT diameter, $d$ [76], [77], [78], as given by $Pos(RBM) = \frac{C_1}{d} + C_2$. Combining Pos(*RBM*), with excitation wavelength and the '*Kataura plot*' [64], [65], it is, in principle, possible to derive the SWNT chirality [79], [80].

Matching the diameter given by Pos(RBM) with excitation wavelength in the Kataura plot also gives information on the s- or m- character. A variety of $C_1$ and $C_2$ were proposed for this relation [75], [76], [77], [80], [81]. Here we use the $C_1$=214.4 cm$^{-1}$ nm and $C_2$=18.7 cm$^{-1}$, from Ref. [76]. These were derived by plotting the resonance energy as a function of inverse RBM frequency without additional assumptions. However, we also validated our results by using the parameters proposed in Refs. [77], [81] and [82]. Raman spectroscopy also probes possible damage via the D peak [83], [84], [85]. The typical Raman spectrum in the 1500-1600 cm$^{-1}$ region consists of the *G$^+$* and *G$^-$* bands. In s-SWNTs, they originate from the longitudinal (LO) and tangential (TO) modes, respectively, derived from the splitting of the $E_{2g}$ phonon of graphene [85]. The positions of the *G$^+$* and *G$^-$* peaks, Pos(*G$^+$*), Pos(*G$^-$*), are diameter dependent and the separation between them increases with decreasing diameter [86], [87]. In m-SWNTs, the assignment of the *G$^+$* and *G$^-$* bands is the opposite, and the full width at half maximum of the *G$^-$* peak, FWHM(*G$^-$*), is larger and Pos(*G$^-$*) down-shifted with respect to the semiconducting counterpart [86], [88]. Thus, a wide, low frequency G- is a fingerprint of m-SWNTs. On the other hand, the absence of such a feature does not necessarily imply that only s-SWNTs are present, but could signify that m-SWNTs are off-resonance.

Doping could also modify positions and FWHMs [89], [90]. In m-SWNTs, a Pos($G^-$) blueshift, accompanied by a FWHM($G^-$) decrease, is observed with electron or hole doping [91], [92]. In s-SWNTs doping upshifts Pos($G^+$), but does not affect FWHM($G^+$) [89], [91].

Thus, a large number of excitation wavelengths are necessary for a complete characterization of SWNTs [78], [81]. Nevertheless, useful information can be derived even with few excitations, especially for process monitoring, when Raman compares the "raw" material with end product.

Raman spectra are taken both on the un-sorted and sorted dispersions deposited on a Si substrate with a Renishaw system at 514 nm (2.41 eV) 633 nm (1.96 eV) and 782 nm (1.58 eV), using a 50X objective and less than 1 mW on the sample. The RBM detection is limited by the cut-off of the notch and edge filters. These are at 120, 100 and 110 cm$^{-1}$ for 514, 633 and 782 nm, respectively, limiting the detection of tubes with diameter up to ~1.9 nm. We use Lorentzians to fit the RBM, and $G^-$ and $G^+$ peaks.

## CNT-based transistors

After separation, the SWNTs are used for CNT-FETs. The devices fabricated from each SWNT samples are based on a spin coating enriched s-SWNTs dispersion to form a network of SWNTs on *n*-type Si substrate (acting as the device *gate*) covered with 250 nm thermally grown SiO$_2$ as gate dielectric. After annealing at 200 °C for 20 minutes, Au electrodes are patterned, via electron beam lithography, to form the *source* and *drain*. The devices are then annealed again to evaporate residual solvents at 200 °C for 5 minutes, and characterised with a *Keithley 4200 Analyzer* in ambient conditions (both pressure and temperature).

## RESULTS AND DISCUSSION

### Room Temperature Separation

The optical absorption data of the four different sets of SWNT before and after the separation process at RT is shown in Figure 2. The absorption peaks are denoted as $M_{11}$, $eh_{11}$(S$_{11}$), $eh_{22}$(S$_{22}$), and $eh_{33}$(S$_{33}$), and assigned to the first transition of m-SWNTs and the first, second and third excitonic transition of s-SWNTs, respectively [64], [65]. The spectrum for the un-sorted DIPS-SWNTs (Figure 2a) shows broad bands for the $M_{11}$, $S_{22}$, and $S_{33}$ peak due to the "nominally" large diameter distribution [93]. Moreover, the lack of sharp features for the excitonic transitions of both m- and s-SWNTs is a signature of bundles [59]. Contrary, the top and bottom bands of the sorted samples eluted through the column show more intense and "sharp" absorption features. The top band has an increased absorption in the $eh_{22}$ and $eh_{33}$ region. These bands are well resolved, unlike the spectrum of the starting material. We estimate, by using the derivative of the absorption, ~76% s-SWNTs, the rest being m-SWNTs, with absorption peaks at ~801 and 865 nm, that we assign [65] to *(13,10)* and/or *(17,10)* and *(17,8)* and/or *(22,1)* tubes.

Better resolved, compared to the DIPS-CNT samples, absorption spectra are obtained for the AD-CNT tubes in Figure 2b. The top fraction shows an increase of the absorption peaks corresponding to $eh_{22}$ in the 800-1100 nm range, with respect to the starting material. We assign [65] the peaks at ~655, ~700 and ~747 nm detected in the $M_{11}$ region, to *(10,10)*, *(13,7)* and/or *(12,9)*, *(19,1)* and/or *(18,3)*, respectively. Derivative of the absorption give ~90% s-SWNTs in this fraction. The spectrum of the bottom fraction has an increased absorption band in the region 550-800 nm, with a broad and not resolved band also detected in the region 800-1200 nm, due to $eh_{22}$.

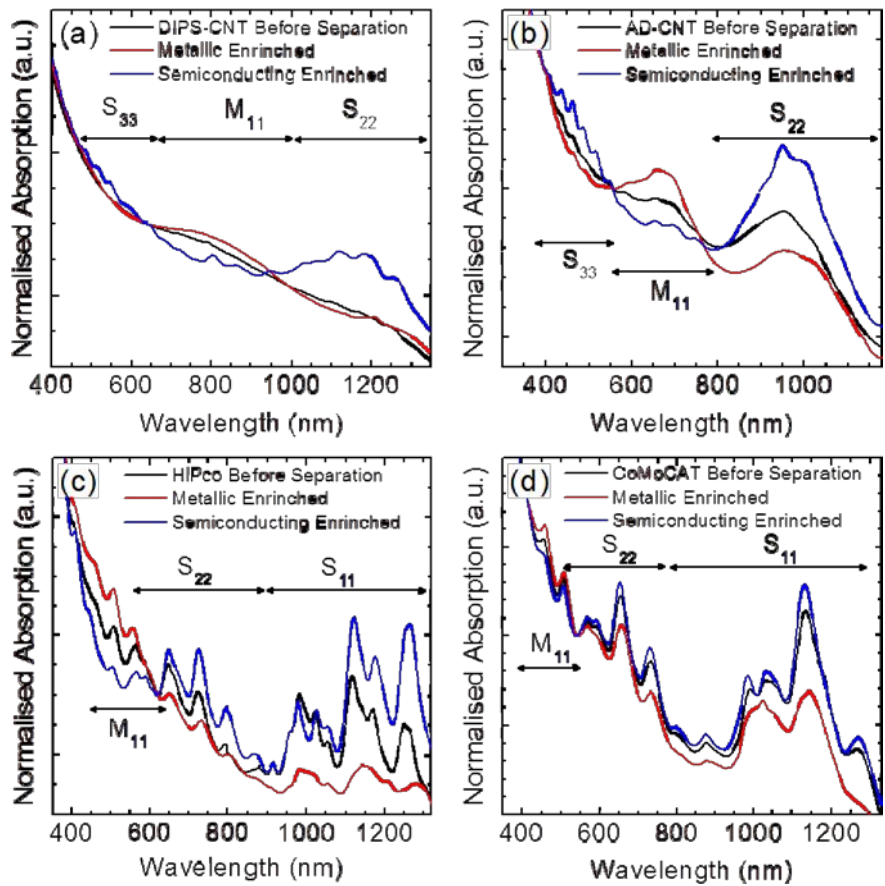

Figure 2: Absorption spectra of SWNTs before and after separation for (a) DIPS-CNT, (b) ADCNT, (c) HiPco, and (d) CoMoCAT. The labels $S_{11}$, $S_{22}$, $S_{33}$ and $M_{11}$ refer to the first, second, third semiconducting and the first metallic excitonic transition, respectively, and are a guide to the eye, since overlap between different excitonic transitions exists[60, 61]. The spectra are normalized for clear visualization.

The OAS of HiPco (Figure 2c) and CoMoCAT (Figure 2d) SWNTs have sharper peaks than the DIPS and ADCNT samples (both for starting material and sorted samples). The top band of sorted HiPco SWNTs (blue curve in Figure 2c) has a reduced absorption in the $M_{11}$ region and an enhancement of the peaks in the $eh_{11}$ region with respect to the starting material (black curve in Figure 2c). The peaks in the region 400-600 nm are assigned [64], [65] to the $eh_{22}$ of *(7,3)*, 505 nm; *(9,2)*, 552 nm; *(6,5)*, 566 nm;

*(8,4)*, 591 nm; *(11,1)*, 611 nm; and $M_{11}$ of *(6,3)* and *(5,5)* at 440 and 411 nm, respectively. There is also an increase in $eh_{11}$ absorption with respect to the starting material. The calculated percentage of s-SWNTs in this fraction is ~90%, with 10% m-SWNTs. The spectrum of the bottom fraction shows an increase in absorption in the region 400-600 nm. We assign [65] the peaks to the following m-SWNTs: *(5,5)*, 411 nm; *(6,4)* and *(7,6)* 455 nm; *(7,7)* and *(9,0)* 507 nm; *(8,8)* and *(11,2)* 558 nm; *(9,9)*, 598 nm; *(10,10)*, 653 nm; *(11,11)*, 733 nm.

However, some of the peaks are quite broad, being the results of a combination of $M_{11}$ and $eh_{22}$ transitions. Indeed, absorption features from s-SWNTs are detected in the $eh_{11}$ region (800-1300 nm), with broad absorption peaks associated to *(8,3)* 952 nm; *(6,5)*, 983 nm; *(7,3)*, 990 nm; *(7,5)*, 1020 nm; *(8,1)*, *(10,2)*, *(8,4)*, *(8,6)* around 1020 nm; other larger diameter SWNTs at ~1150 nm [64], [65]. The estimated percentage of s- and m-SWNTs in the bottom fraction is ~38% and ~62%, respectively.

Similar results to the HiPco one are obtained with the CoMoCAT SWNTs (Figure 2d). The top band has a reduced $M_{11}$ absorption and an enhancement in the $eh_{11}$ and $eh_{22}$ regions with respect to the starting material. In particular, there is a reduction of *(6,6)* absorption with respect to the un-sorted sample. The bottom band shows an increase of absorption in the region 400-550 nm, with peaks assigned to *(6,6)* and *(7,4)* at 458 nm; *(7,7)* and *(9,0)* at 507 nm; *(9,9)* at 598 nm; *(10,10)* at 658 nm; *(11,11)* at 733 nm [65]. However, these bands are quite broad due to the combination of $M_{11}$ and $eh_{22}$. s-SWNTs are detected via the $eh_{11}$ absorption in the region 8001300 nm. We estimate that in the top fraction there are ~78% s-SWNTs, while in the bottom fraction there is an enrichment of m-SWNTs of ~65%.

The Raman spectra of un-sorted and sorted samples are analyzed and compared in Figures 3-6. Figures 3(a,b) plot the spectra at 514.5 nm for DIPS. This wavelength is expected to be in resonance mostly with the third excitonic transition of s-SWNT [64], [65]. The spectrum in the RBM region of the starting material shows features related to m- and s-SWNTs [76]. In particular RBMs of s-SWNTs *(16,6)*, *(13,8)*, *(14,7)*, *(13,5)*, *(15,2)* and m-SWNTs *(8,5)* and *(9,3)* are detected. After sorting, we still detect the RBMs of s-SWNTs in the top band. On the contrary, in the bottom band we detect only RBM assigned to *(8,5)* and *(9,3)*. In the *G* peak region (Figure 3b), the un-sorted sample shows the typical $G^+$ and $G^-$ peaks expected from a mixture of s-and m-SWNTs [76], [87], [94]. FWHM($G^-$) of the top band is ~47% narrower with respect to the un-sorted sample. FWHM($G^-$) of the bottom band is ~28 and ~89% larger with respect to the un-sorted sample and the top band, respectively. The value of FWHM($G^-$) is an indication of higher s- and m-SWNTs content in the top and bottom band, respectively. Doping could also modify the peaks' shape [89], [91]. We do not expect doping to play a major role here for two reasons. First, both samples contain the same surfactants, even if with different concentration, so the doping contribution should be similar for both. Second, Pos($G^-$) should upshift and become narrower as a consequence of both p or n doping [89], [91].

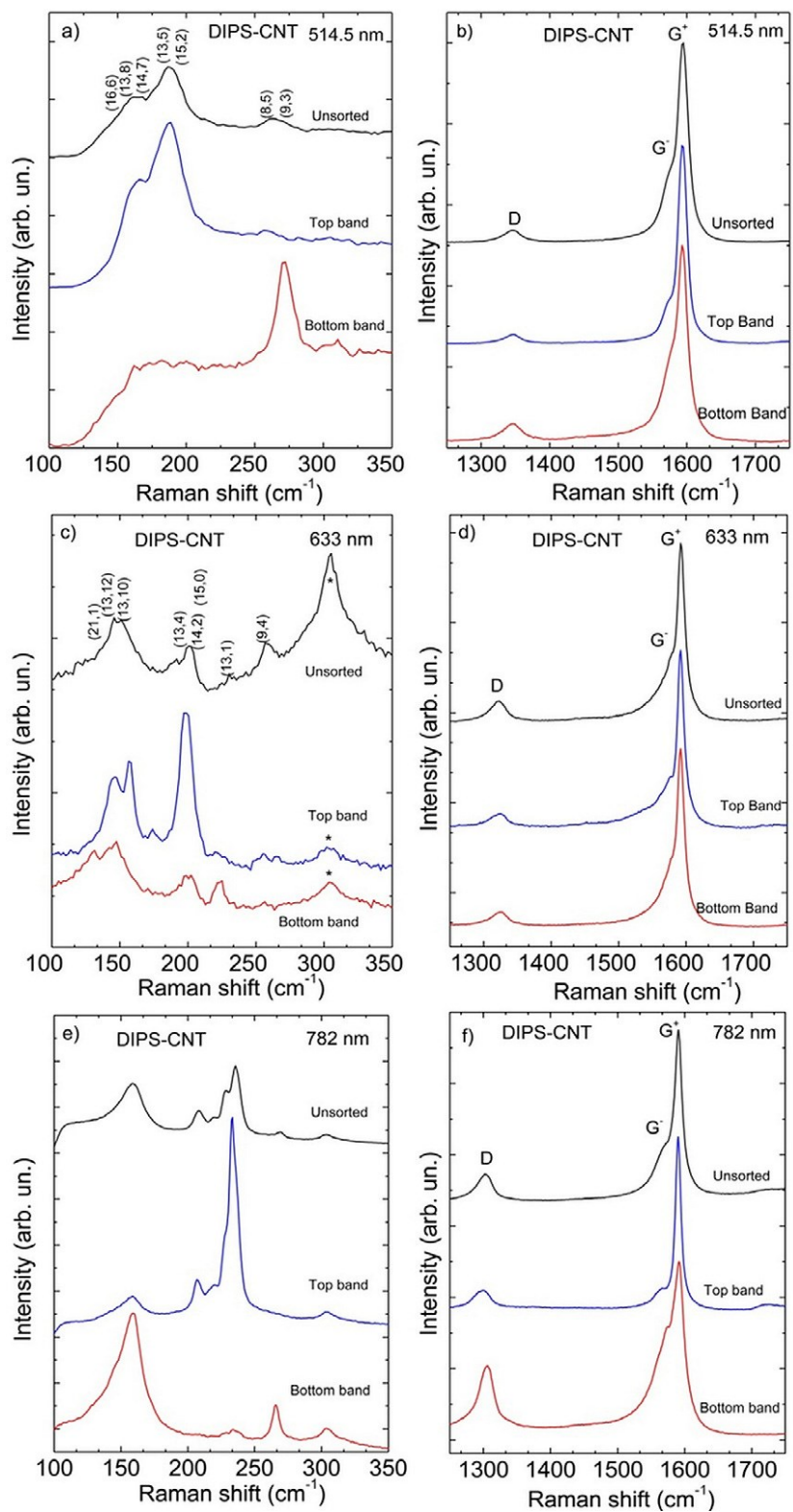

**Figure 3:** Raman spectra of DIPS SWNTs. (a) RBM and (b) G region measured at 514.5 nm, (c) RBM and (d) G region measured at 633 nm and (e) RBM and (f) G region measured at 782 nm. A peak from the Si substrate is marked with * in (c).

Figures 3(c, d) plot the spectra at 633 nm. The RBM region (Figure 3c) shows an enhancement of the peak associated with the *(13,4)* s-SWNT in the top fraction. This peak is reduced in intensity in the bottom fraction. In the same fraction there is also an increase in intensity of the peaks assigned to *(14,2)*, *(15,0)* and *(13,1)*. The FWHM($G^-$) of the top fraction (23.4 cm$^{-1}$, Figure 3d) is ~70% narrower with respect to the bottom band, which could indicate a higher percentage of s-SWNTs.

Evidence of MS-separation is seen also at 782 nm. While the RBM region (Fig.3e) of the un-sorted materials show peaks associated both to s- and m-SWNTs, in the top fraction there is a decrease in intensity of peaks at ~160 cm$^{-1}$, associated to *(19,1)* and *(18,3)*, and an enhancement of peaks in the region 200-235 cm$^{-1}$, due to *(14,1)*, *(9,7)*, *(8,7)*, *(11,3)* and *(12,1)* [65], [76]. The latter peaks observed both in the starting material and the top band, have very low intensity in the bottom fraction.

Figure 3e also indicates the presence of m-SWNTs such as *(19,1)* and *(18,3)*. In the *G* peak region (Figure 3f), the FWHM($G^-$) of the top band is ~66% narrower with respect to the unsorted sample. While FWHM($G^-$) for the top band is ~26% and ~106% larger with respect to the un-sorted sample and the top band, respectively. This indicates higher s- and m-SWNTs content in the top and bottom band, respectively.

Figures 4(a, b) plot the spectra at 514.5 nm for AD-CNTs. The spectra are not significantly different. This is due to the resonant excitation of s-SWNTs at this wavelength [65], [76]. The Raman results agree with the absorption data shown in Figure 2b. Indeed, the spectrum of the bottom band has FWHM($G^-$) only ~3 cm$^{-1}$ larger than the top, too small to be statistically significant of a higher m-content. Similar spectra are also measured at 633 nm. However, at this wavelength, mostly m-SWNTs are in resonance. The RBM region (Figure 4c) of the bottom fraction shows an enhancement of the peaks at ~200 cm$^{-1}$ (assigned to m-SWNTs) with respect to the un-sorted sample and the top band, respectively. In the G peak region (Figure 4d), the FWHM($G^-$) of the top band sample is ~35% narrower than the un-sorted sample and the bottom band. The Pos($G^-$) is also ~10 cm$^{-1}$ (1559 cm$^{-1}$) upshifted with respect to the un-sorted sample (1550 cm$^{-1}$) and bottom band (1548 cm$^{-1}$). Clear evidence of MS-separation is seen at 782 nm. The RBM region (Figure 4e) in the top fraction shows an intensity reduction of the peaks ~165 cm$^{-1}$, due to *(15,6)*, and *(12,8)* [65], [76], while the peaks at ~210 cm$^{-1}$ and ~238 cm$^{-1}$ due to s-SWNTs are enhanced in the top band and decrease (~210 cm$^{-1}$) or disappear (~238 cm$^{-1}$) in the bottom band. FWHM($G^-$) is broader (~24%) in the bottom band, with respect to both the unsorted sample and the top band, fingerprint of m- separation, see Figure 4f.

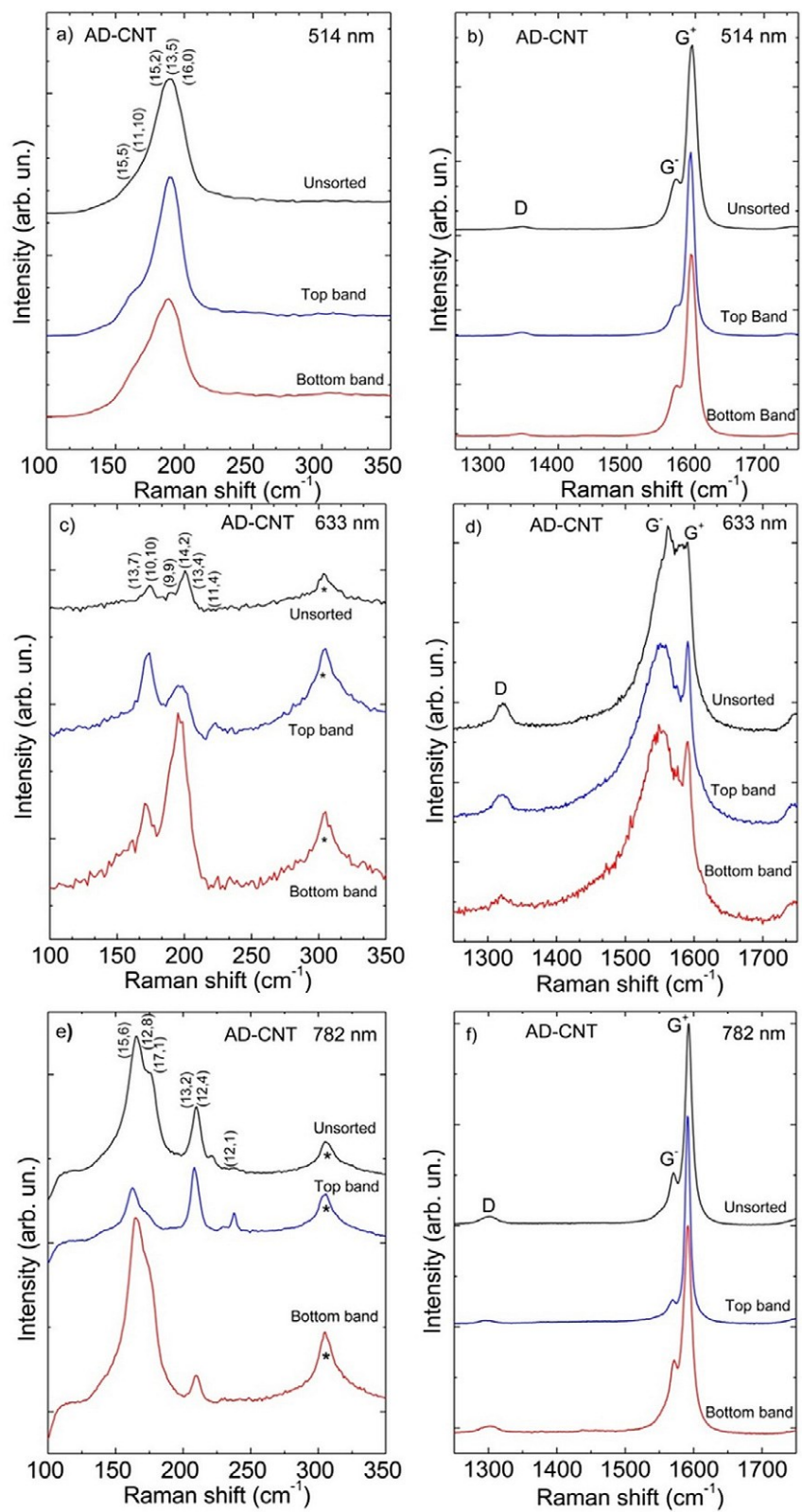

**Figure 4**: Raman spectra of AD-CNT SWNTs. (a) RBM and (b) G region measured at 514.5 nm, (c) RBM and (d) G region measured at 633 nm and (e) RBM and (f) G region measured at 782 nm. A peak from the Si substrate is marked with *.

The Raman spectra for HiPco samples are reported in Figure 5. The spectra in the RBM region at 514.5 nm after enrichment are rather different with respect to the un-sorted samples (Figure 5a). The top band show peaks in the range 170-215 cm$^{-1}$, while these have a reduced intensity in the bottom band.

These peaks are mainly assigned to s-SWNTs: *(11,9)*, *(16,2)*, *(10,9)* and *(14,3)*. The top band shows a decrease in intensity of the peaks in the range 250-320 cm$^{-1}$. In particular, no peak is detected at ~316 cm$^{-1}$, assigned to *(8,2)* [65], [76], in the top fraction, while this is enhanced with respect to the un-sorted sample in the bottom band.

In Figure 5b, the FWHM($G^-$) of the top band is ~60% narrower with respect to the un-sorted sample and the bottom band. Moreover, Pos($G^-$) is ~22 cm$^{-1}$ upshifted with respect to the bottom band. This is an indication that the contribution of Pos($G^-$) in the top band is the combination of m- and s-SWNTs [91], the latter having a diameter~1.3 nm [86]. The bottom band has Pos($G^-$) ~1543 cm$^{-1}$ with main contributions from m-SWNTs with 0.7-0.9 nm diameter [86]. Figures 5(c, d) show Raman spectra measured at 633 nm. At this excitation both m- and s-SWNTs are expected to be in resonance [65]. From Figure 5c, m-SWNTs are detected in the 150-225 cm$^{-1}$ range, while s-SWNTs in the range 240-350 cm$^{-1}$. The spectra of the bottom band show an increase in the RBMs intensity at~150-225 cm$^{-1}$, in particular at~190 cm$^{-1}$, assigned to *(12,3)* [65], [76], and a reduction of the RBM peaks in the 240-350 cm$^{-1}$ region with respect to the unsorted sample and the top band. The latter shows the opposite behavior, with an enhancement of the peaks assigned to s-SWNTs. In the *G* peak region (Figure 5d), evidence of splitting of the *G*$^-$ peak is seen in all spectra. The peak at lower wavenumber (~1542 cm$^{-1}$), due to LO mode of metallic tubes with diameter ~1.0-1.1 nm [86], has the same position in the three spectra, but the FWHM is ~63% larger in the spectrum of the bottom band with respect to the starting material. The peak at ~1555 cm$^{-1}$, due to a combination of LO of m-SWNTs with diameter of ~1.2 nm and TO of s-SWNTs with diameter~1.0 nm [86], has both the same position and FWHM in the three spectra. FWHM($G^+$) is reduced both in the top (63%) and bottom (32%) fractions with respect to the starting material.

Figures 5(e, f) report the spectra in the RBM and G peak region, respectively, measured at 782 nm for the HiPco samples. This excitation wavelength is in resonance mainly with the s-SWNTs eh$_{22}$ [65]. The RBM region shows an enhancement of the peaks in the spectral range 200-250 cm$^{-1}$ for the top fraction, with respect to the un-sorted sample, while the same peaks have a reduced intensity in the bottom fraction. The detection of mainly s-SWNTs in all three samples is also evidenced by the *G* peak region. Pos($G^+$) and Pos($G^-$) have are~1590 and ~1559 cm$^{-1}$ for the three samples. Moreover, FWHM($G^-$) has a similar width (~15 cm$^{-1}$) in the top and bottom bands.

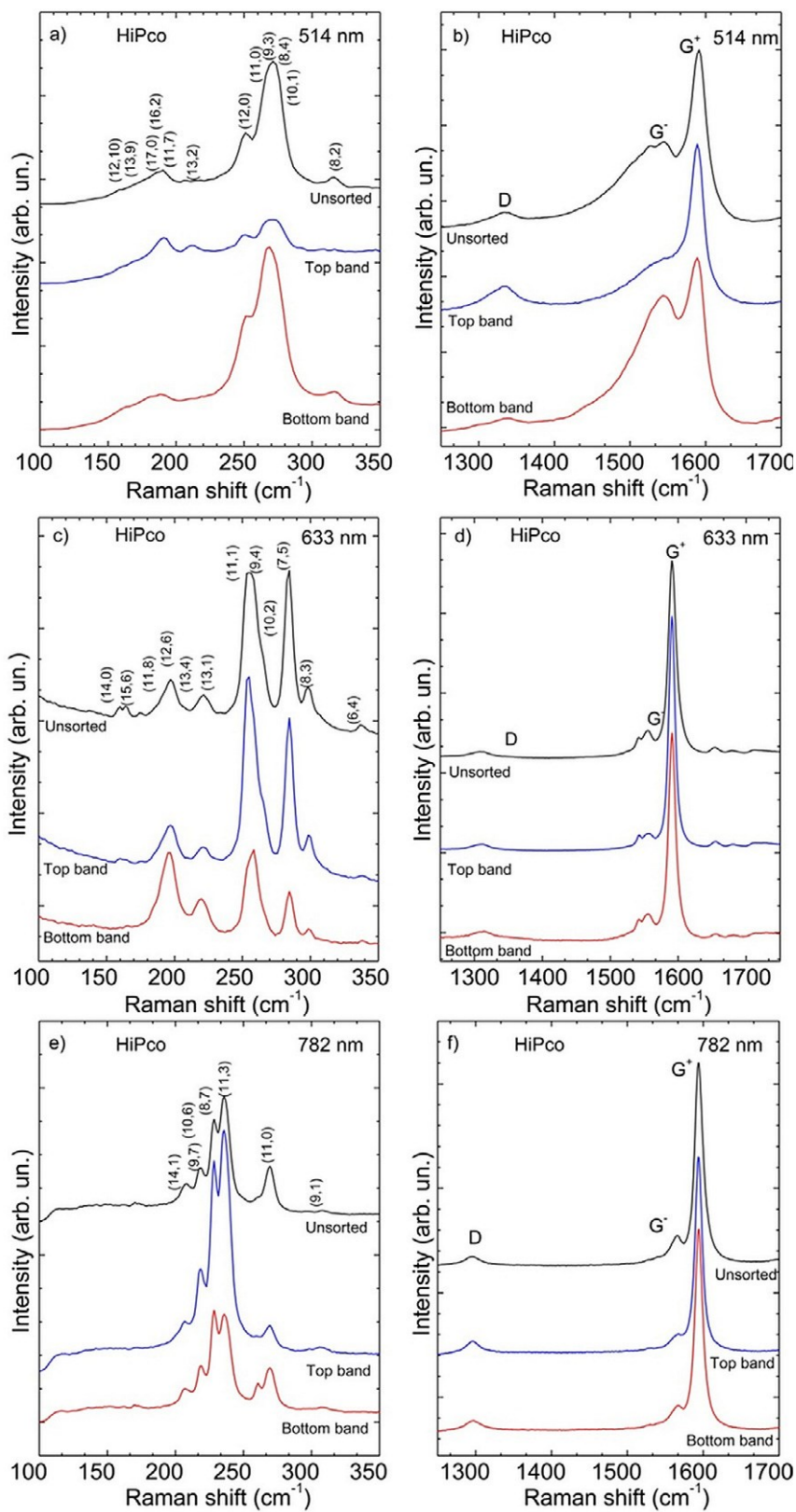

**Figure 5:** Raman spectra of HiPco SWNTs. (a) RBM and (b) G region measured at 514.5 nm, (c) RBM and (d) G region at 633 nm and (e) RBM and (f) G region measured at 782 nm.

As for the HiPco samples, also in CoMoCAT 514.5 nm excitation probes both m- and s-SWNTs [65]. The RBM spectra are reported in Figure 6a. The spectra show two distinct regions: 150-200 $cm^{-1}$, assigned predominantly to m-SWNTs, and the other (with more intense peaks) in the 225-320 cm$^{-1}$ range, associated with resonance of both m- and s-SWNTs [65]. The spectra of the top band have lower intensity than the bottom one. There is also a signal associated to s-SWNTs (*i.e. (8,4)*). This chirality along with the *(7,6)*, *(10,2)* tubes are also detected in the absorption spectrum of the bottom band, Figure 2d. The higher content of s- and m-SWNTs in the top and bottom fractions, respectively, with respect to the un-sorted sample is evidenced by the analysis of the *G* peak (Figure 6b).

Indeed, FWHM($G^-$) of the top band is ~54% narrower and Pos($G^-$) is ~21 cm$^{-1}$ upshifted with respect to the un-sorted sample, while we detect a ~23% widening of FWHM($G^-$) and ~13 cm$^{-1}$ downshift of Pos($G^-$) with respect to the un-sorted sample. At 633 and 782 nm, mostly s-SWNTs are in resonance and no clear separation is detected from the Raman analysis. Absorption measurements in Figure 2d indicate s-SWNTs (*i.e. (7,5)*, *(7,6)*, *(10,2)*, *(9,4)*, *(8,4)* etc.) [64], [65], in the bottom band as well. These tubes are also detected in the RBM region (see Figure 6c). The excitation of mainly s-SWNTs at 633 nm is also clear from the analysis of the *G* peak region (Figure 6d). Pos($G^+$), Pos($G^-$) and FWHM($G^-$), FWHM($G^+$) have similar values for the three samples. The same holds also for 782 nm excitation, where s-SWNTs are in resonance for CoMoCAT samples. The RBM region (Figure 6e) shows the presence of s-SWNTs in all three samples, in agreement with absorption results, where an enrichment of s-SWNTs is detected in the top fraction (~78%), while the bottom fraction, although enriched in m-SWNTs, still contains ~35% s-SWNTs. FWHM($G^+$) for the top band is narrower compared to the unsorted one (11 instead of 14 cm$^{-1}$), see Figure 6f. However, this reduction is not sufficient to indicate a narrower diameter distribution. Due to the diameter dependence of Pos($G^+$) [86], [87], the removal of tubes with large difference in diameter, will reduce FWHM($G^+$) [86], [87]. Pos(G) is 3 cm$^{-1}$ downshifted and FWHM($G^-$) ~47% larger in the bottom with respect to the top band. To summarize, Figures 3-6 show that after sorting using gel agarose filtration, we have an enrichment of s- and m-SWNTs in the top and bottom bands, respectively. The separation is more effective for s- than m-SWNTs, in agreement with the absorption spectra in Figure 2.

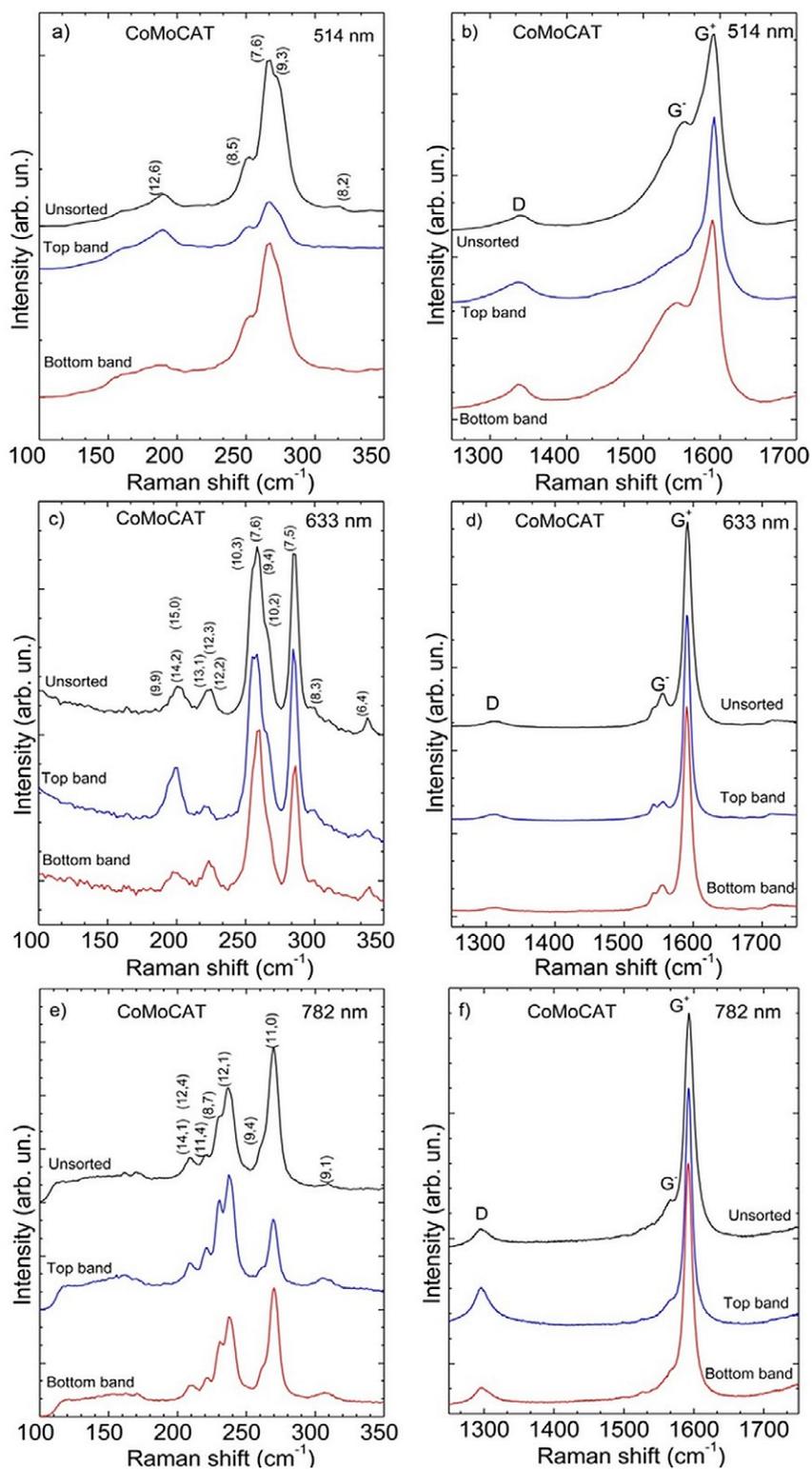

**Figure 6:** Raman spectra of CoMoCAT SWNTs. (a) RBM and (b) G region measured at 514.5 nm, (c) RBM and (d) G region at 633 nm and (e) RBM and (f) G region measured at 782 nm.

*Temperature dependence*

To understand the temperature influence on the sorting process we carry out the gel agarose beads filtration process in a range of temperature between 6 and 50 °C, Figure 7. The temperature limits were chosen based on the instability of SDS and agarose gel below 6 °C and above 50 °C, respectively. For AD-CNT, HiPco and CoMoCAT, the separation process at RT shows distinct coloration between the top and bottom band [33], [37], [95]. For DIPS-CNT, there is no clear color distinction between the top and bottom bands, they both are black or grey. This is due to the large tube diameter distribution that causes the different colors to overlap. The colors of SWNTs with different chirality depend on the optical absorption peak [39]. The colors of sorted SWNTs in Figure 7 change according to the separation temperature. This could be ascribed to the presence of diverse SWNT species, due to the sorting process, that absorb at different wavelengths.

For different separation temperature, the variation in fraction concentration between the top and bottom band is observable, indicating a difference between the quantities of tubes in respective bands after separation. Their relative concentrations can be determined from the optical absorption analysis presented in the supporting information (Figure S5). Generally, the bottom bands are more concentrated for low temperature process, and the top bands are more concentrated at high process temperature. The exact relative concentration is dependent on the SWCNTs source. From Figure S5, the maximum bottom band to top band concentration ratio is ~10 and ~0.1 at 6 °C and 50 °C, respectively.

The effectiveness of the separation of m- and s-SWNTs changes with process temperature both for the top and bottom bands, as can be seen from the OAS in Figure 8. The bottom band (Figure 8a) extracted for DIPS-CNTs shows the highest enrichment in m-SWNTs at RT, while the top band (Figure 8b) shows the highest (~80%) percentage of s-SWNTs at 10 °C. A similar trend is also observed for AD-CNTs. The bottom band for AD-CNTs (Figure 8c) has the highest percentage of m-SWNTs at 30 °C, while the top bands (Figure 8d) has the highest percentage of s-SWNTs (~95%) at 6 °C. The same trend, with a better enrichment of m-SWNTs in the bottom bands at RT or slightly higher, and s-SWNTs in the top bands at low temperatures is also valid for HiPco samples, Figures 9(a,b). As for AD-CNTs, HiPco nanotubes show a higher enrichment (~70%) of m-SWNTs in the bottom fractions, Figure 9a, at 40 °C, while the s-SWNTs in the top fractions are more (~95%) at 6 °C, see Figure 9b.

The effect of temperature for CoMoCAT tubes is less pronounced with respect to the other samples. However, Figures 9(c, d) show higher m-SWNT enrichment in the bottom band at 40 °C (Figure 9c) while, contrary to the other tube sources, the top band (Figure 9d) shows higher enrichment of s-SWNTs at RT.

The highest separation purity obtained at optimum temperatures for s-SWCNTs (6 °C) and m-SWCNTs (40 °C) are ~95% and ~70%, respectively. Whilst the highest s-SWCNT purity obtained here (~95%) is comparable to figure reported in ref. [46] using the same separation method, the purity for m-SWCNT recorded here (~70%) is considerably less. However, The presented purity figure for m-SWCNT is conservatively chosen based on repetition of the process. The study of temperature here has shown it to be an important parameter in determining the purity percentage of the separation, a detail previously not discussed. Consequently, depending on the electronic type required, i.e. s-SWCNT or m-SWCNT, one has to optimize the temperature parameter accordingly. It gives additional control to any already optimized parameters. Furthermore, we have shown that the correlation between process temperature and separation purity is consistent for all four different SWCNT sources and we believe that any separation carried out at the optimized respective temperatures would yield superior purity compared to the process at room temperature.

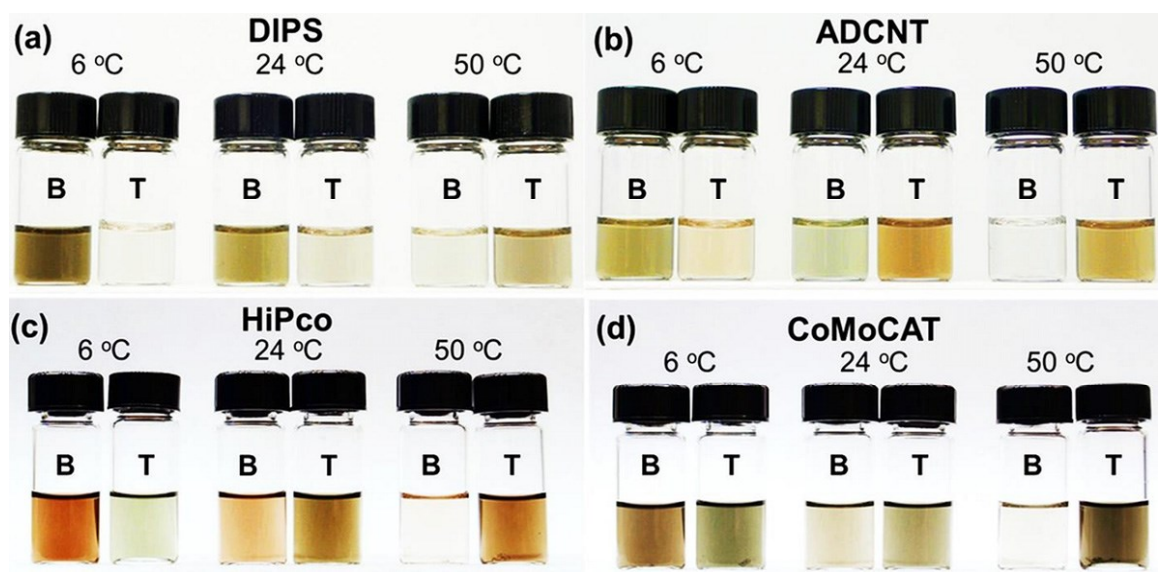

**Figure 7:** Bottom (B) and top (T) band fractions for three different process temperatures. (a) DIPS-SWNT, (b) AD-CNT, (c) HiPco, (d) CoMoCAT. The colours of the dispersions vary moving from 6 to 50 °C and for different SWNT sources.

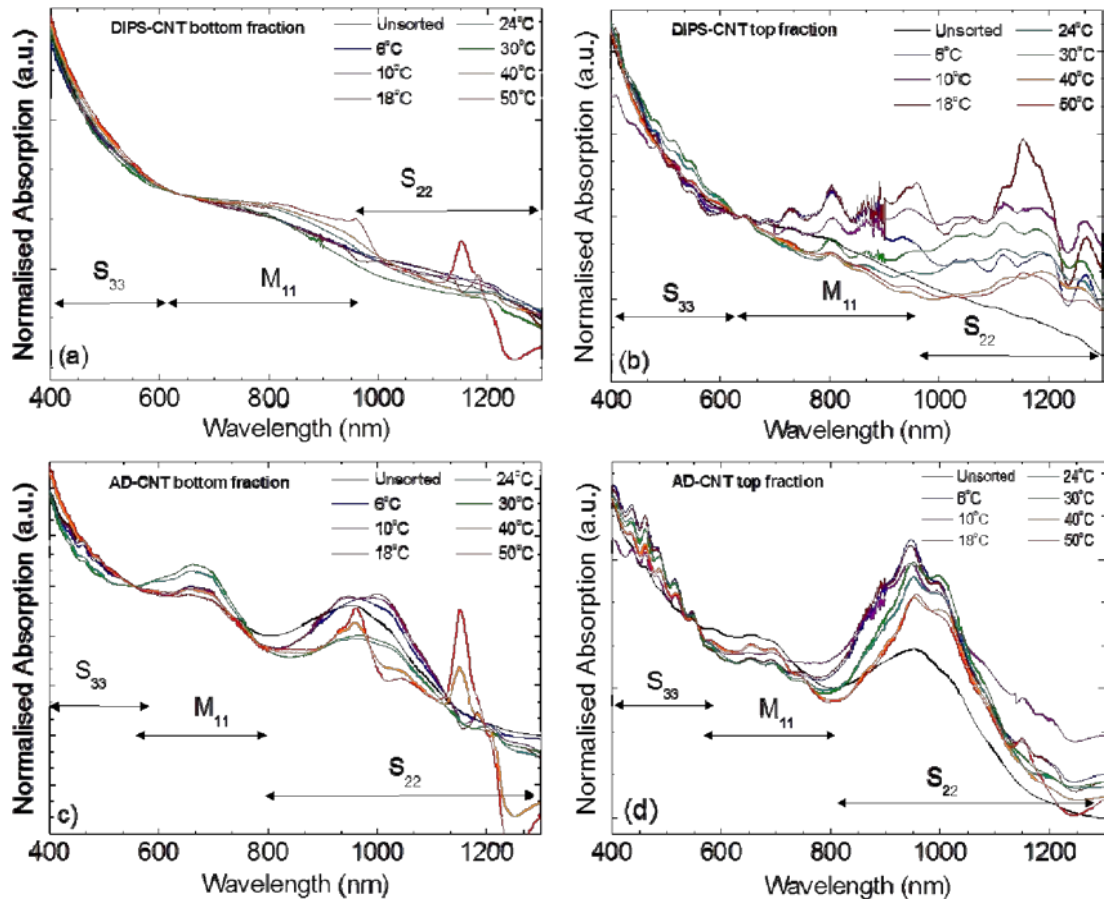

**Figure 8:** Optical absorption spectra for DIPS-CNTs (a) bottom and (b) top band, AD-CNTs (c) bottom and (d) top band. The peaks between 950 nm and 1150 nm in (b) and (d) are instrument originated noise associated with changing of source and detector.

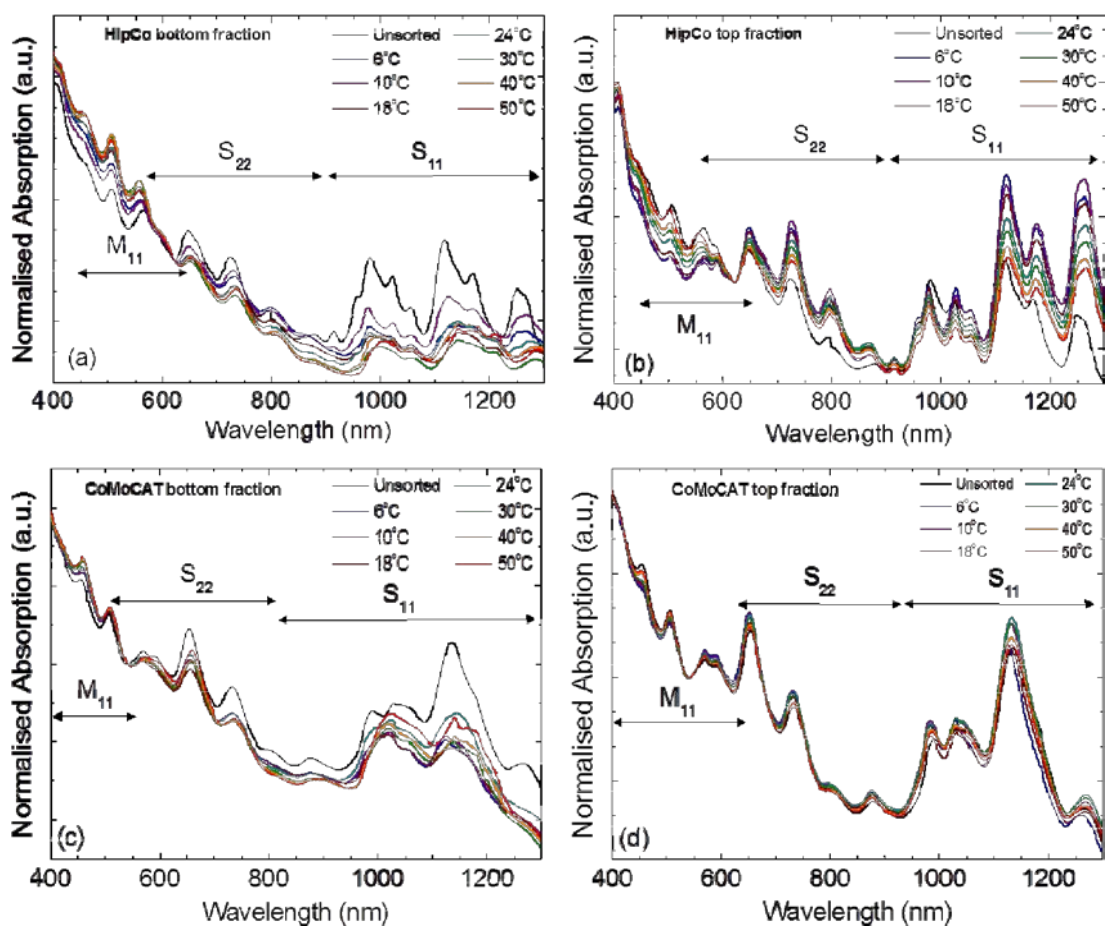

**Figure 9:** Optical absorption spectra for HiPco (a) bottom and (b) top band and CoMoCAT (c) bottom and (d) top band.

The Raman spectra of the sorted fractions for all the four sources are reported in supplementary information. The Raman analysis confirms the data from OAS, showing higher enrichment of s-SWNTs at low temperatures and m-SWNTs at temperatures higher than RT.

PLE spectra measurements were carried out to analyze the content of s-SWNTs in the sorted fractions and how their composition changes with temperature. Due to the diameter range of DIPS-CNT and AD-CNT samples, the PL emission mostly fall in the infra-red region and therefore cannot be measured by our detector. Figure 10 shows the PLE contour maps of sorted HiPco and CoMoCAT, at 6, 24 and 50 °C. The PLE maps for the un-sorted materials are reported as well. The emission intensity contour of these maps are normalized using the same factor used for the OAS. Figures 10,11 show that the PLE signals are lower in the bottom fractions. Also, the PLE intensity in the bottom fractions decreases as temperature increases.

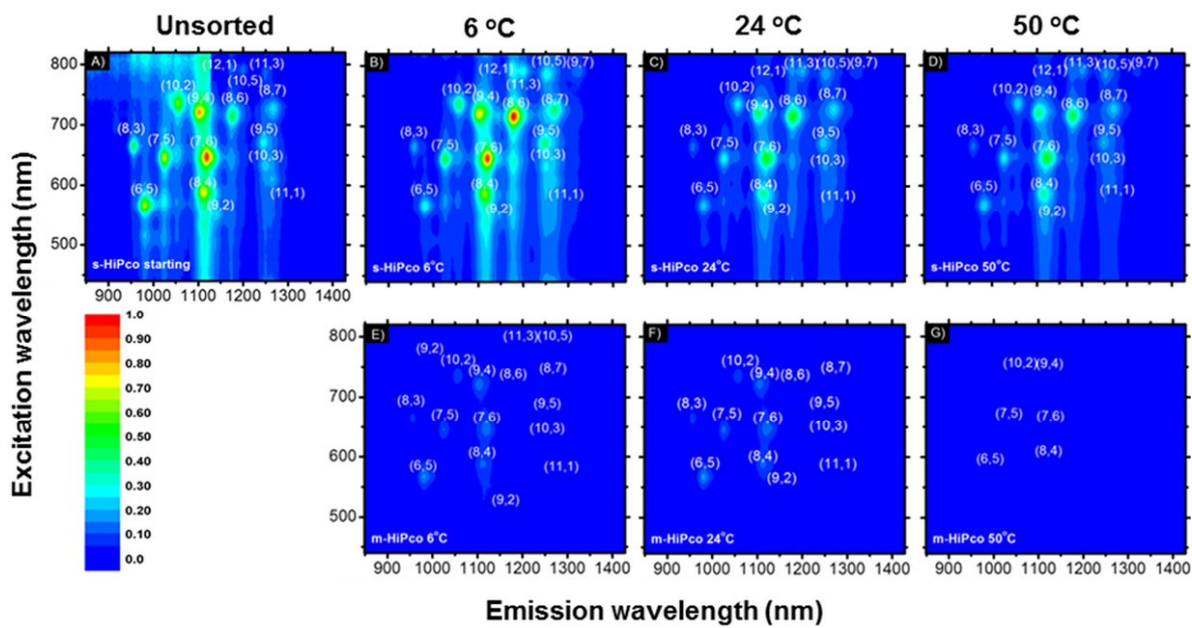

**Figure 10**: PLE maps of a) un-sorted and sorted fractions of HiPco at 6, 24 and 50 °C, for b-d) top and e-g) bottom fractions. The emission intensity contour are normalised with the same factor used for the OAS in Figure 9a.

Figures 10,11 are PLE maps of the starting materials and the sorted sample dispersions for HiPco and CoMoCAT, respectively. Figure 10a is a PLE map of un-sorted HiPco. PL from excitonexciton resonances of tubes with different chiralities, phonon sidebands, and EET [57], [58] are detected. The strongest EET ($eh_{22}^D, eh_{11}^A$) features, *i.e.* from *(6,5)* to *(7,5)*, from *(7,5)* to *(10,2)* and from *(7,6)* to *(8,6)* have PL emission comparable to the ($eh_{22}, eh_{11}$) emission of other chiralities.

The EET features are an indication of the presence of bundles [57], [58]. Figure 10b shows a decrease in PL emission from smaller diameter tubes, such as *(6,5)*, *(8,3)* and *(7,5)* in the top fraction of the sorted sample at 6 °C with respect to the un-sorted dispersion. An increase in PL emission from larger tubes, such as *(8,6)*, *(8,7)*, *(9,5)*, etc. is detected in the same sample with respect to the un-sorted dispersion. The PLE intensity of the sorted top fraction of HiPco SWNTs at 24 and 50 °C (Figures 10c, d), decrease with temperature. EET features are also detected in these PLE maps. Figures 10 (e-g) plot the PLE maps of the bottom fraction of sorted HiPco SWNTs at 6 °C, 24 °C and 50 °C, respectively. The absorption and Raman measurements indicated that these fractions are enriched in m-SWNTs. However, we still detect PL emission from these dispersions, although with low intensity with respect to both the un-sorted dispersion and the top fractions (enriched in s-SWNTs). E.g., the *(7,5)* PL emission has ~20% and ~10% (Figures 10 (e,f), respectively) of the intensity compared to that from same chirality in the unsorted dispersion. Unlike the top fraction (s-SWNTs enriched), the bottom fraction has more m-SWNTs for higher temperature. While at 6 and 24 °C we still detect PL emission from a number of chiralities, the bottom fraction sorted at 50 °C does not have any PL features. These two observations are in agreement with Raman and OAS measurements.

The PLE maps of CoMoCAT samples are reported in Figure 11. As for the HiPco samples, the un-sorted CoMoCAT dispersions also show bundles [57], [58]. Indeed, strong EET features are revealed together with the exciton-exciton resonances of tubes with different chiralities and phonon sidebands. Moreover the PLE maps of the top fraction (Figures 11b-d) processed at different temperatures show a decrease in PL emission from smaller diameter tubes, such as *(6,5)*, *(8,3)* and *(7,5)* in the top fraction of the sorted sample at 6 °C with respect to the un-sorted dispersion.

An increase in PL emission from larger tubes, such as *(7,6)*, *(8,4)*, *(9,4)*, etc. is detected in the same samples with respect to the un-sorted dispersion. The PL emission is educed with increasing process temperature, as shown in the PLE maps of the sorted top fraction of CoMoCAT at 24 °C and 50 °C (Figures 11c, d). EET features are also detected in these PLE maps. The PLE maps of the bottom fraction, for the separation process carried out at 6, 24 and 50 °C, show a low intensity PL emission, Figures 10e-g. As for HiPco samples, the PL intensity decreases with increasing process temperature. However, unlike the HiPco sample (Figure 10g) where no PL emission was detected at 50 °C, for CoMoCAT, a low intensity PL is still detected from *(6,5)*, *(7,5)*, *(8,5)* and *(7,6)*, Figure 11g. This confirms the data from OAS and Raman, pointing to a more effective MS-separation with HiPco compared to CoMoCAT.

The optical characterization shows that there is a correlation between the percentage of m- or s-SWNTs and temperature. The top fractions have increasing s-SWNTs content with decreasing temperature (up to 95% at 6 °C for HiPco SWNTs). The m-SWNT purity in the bottom fraction increases with increasing temperature. Ref. [48] reported that the affinity or interaction strength between s-SWNTs and the agarose gel surface increases with decreasing tube diameter. In the case of DIPS-CNT and AD-CNT, having larger average diameter with respect to CoMoCAT and HiPco, the weaker agarose-SWNT affinity results in more s-SWNT getting eluted along with m-SWNT when the 1% SDS solution is poured into the separation column, resulting in the bottom band having lower percentage of m-SWNTs compared to that of HiPco and CoMoCAT (~70%).

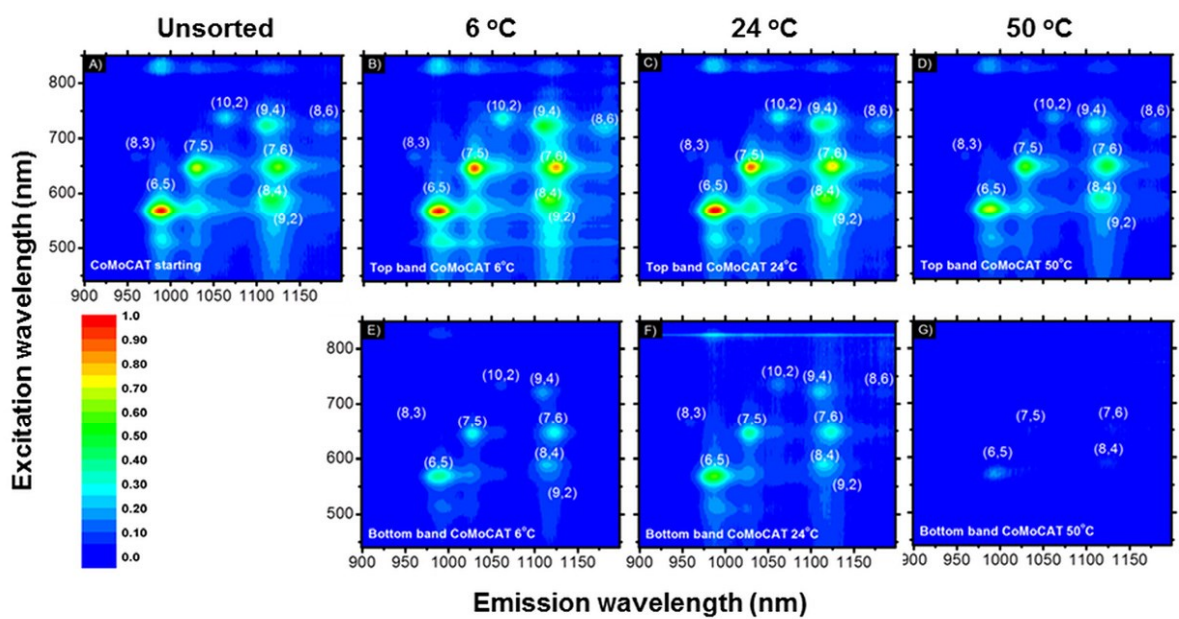

Figure 11: PLE maps of a) un-sorted and sorted fractions of CoMoCAT at 6, 24 and 50 °C, for b-d) top and e-g) bottom fractions. The emission intensities are normalised with the same factor used for OAS in Fig. 9b.

*Separation mechanism*

The separation mechanism is connected with the wrapping and encapsulation [31] of the SDS surfactant molecules around SWNTs. SDS molecules discriminate between SWNTs of different electronic structure by forming different assembly micelles structures on s- and m-SWNTs [96], and the interaction is strongly dependent on pH condition [97] and concentration of SDS molecules [49], [98]. Ion-dipole forces were proposed as the primary interaction mechanism responsible for SWNTs adsorption on agarose gel [99]. Ref. [100] suggested that, due to the electrostatic properties of SWNTs, SDS molecules form flatly assembled and randomly oriented structures around s-SWNTs, and cylindrical micelles around m-SWNTs and concluded that this difference in encapsulation orientation is responsible for the separation process.

The agarose gel surface was shown to have very low physical affinity towards SDS molecules [33]. s-SWNTs with random flat SDS shell have less surfactant coverage. This is responsible for the ineffectiveness of the shielding between s-SWNTs and the agarose gel, driven by the stronger affinity of s-SWNTs towards agarose gel [100]. In contrast, m-SWNTs with ordered cylindrical micelles have high density of surfactants around the surface, forming a brushlike structure [100], [101]. The latter results in a steric hindrance between SWNTs sidewalls and agarose surface. Figure 12 schematically represents this process. Upon flushing 1% SDS solution into the agarose-SWNTs mixture the m-SWNTs are eluted, while the s-SWNTs remain physisorbed onto the agarose surface. By adding a different surfactant aqueous solution, such as SDC, SC, SDBS, Tween 20 and TX-100, all with bigger molecular structures compared to SDS [102], the s-SWNTs can then be removed from the agarose surface and eluted through the separation column [46]. The added surfactants substitute and displace the SDS in the agarose gel column.

Our results are in agreement with the proposed separation mechanism described in ref. [100]. In our case, the adsorption of SDC molecules with the subsequent formation of secondary micelles [31] enhances the steric hindrance between agarose and SWNTs. The s-SWNTs are then eluted following the same process as the metallic counterpart. This proposed separation mechanism could explain why the percentage of m-SWNT in the bottom band is lower than the percentage of s-SWNT in the top band (70% vs. 95%) as reported in Sect. 3.1.

In addition, our data shows evidence that by varying the separation process temperature, the relative size and distribution of the micelles structures around the SWCNTs are affected. Consequently, the resulting separation purity is also affected and dependent on the process temperature variation, for a given agarose concentration (2% agarose). The purity of s-SWNTs in the top band increases with decreasing temperature, while m-SWNTs show the opposite behavior.

Ref. [103] reported a correlation between the purity of the separated SWNTs and the concentration of agarose in the gel matrix, which is related to the pore size. It was suggested that the interaction between SWNTs and agarose gel occurs within the pores in the gel matrix, leading to the separation of SWNTs by electronic type [103]. However, from our study, we suggest that the interaction between SWNTs and agarose gel occurs primarily around the surface of the gel beads [46], rather than within the gel matrix as described in Ref. [103]. Here, the agarose gel beads column acts as a sieve through which the SWNT solution is filtered by means of gravitational pull alone, similar to gel column chromatography [46]. The SWNTs are not embedded in the agarose gel matrix, in which case electrophoresis [27] or gel compression [103] must be employed to separate and release them. Consequently, the variance in separation purity with temperature here may be attributed to the surfactant micelles structure and density around the tubes rather than the change of gel agarose morphology or pore size.

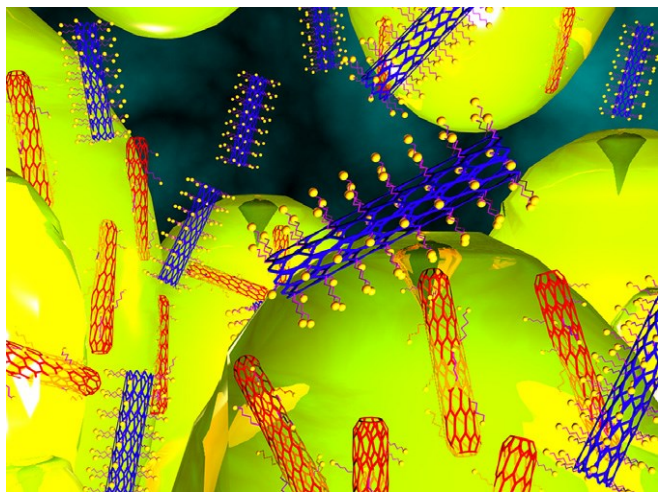

**Figure 12:** Scheme of MS-separation process. Blue tubes are m-SWNTs encapsulated with SDS forming cylindrical micelles; red tubes are s-SWNTs with low concentration of random flat SDS adsorbed on their sidewalls; textured green mass are agarose gel microbeads. The m-SWNTs remain mobile and suspended in the solvent while the s-SWNTs are physisorbed on the surface of the agarose gel microbeads.

There is still a considerable debate regarding the wrapping mechanism and orientation of surfactant molecules around SWNTs. The selectivity of SDS towards m- or s-SWNTs has not been fully clarified yet. Both theoretical [101], [104], [105] and experimental [100], [102], [106] investigations concur that surfactants can have different orientations around SWNTs depending on the surfactant concentration [101] and type [102], as well as tube diameter [101], [105] and electronic type [100]. SDS can either lie randomly on the surface of SWNTs [100], [105], or can form cylindrical micelles perpendicular to the SWNT circumference with the hydrophobic tail attached to the SWNT surface, and the hydrophilic head pointing outwards [100], [101], [102]. It was also shown that SDS micelles are affected by temperature; when the temperature is increased, the SDS micelles' size reduces, the micellar shape becomes distorted and the aggregation number reduces [107].

Therefore, we propose that the separation process temperature affects the distribution of SDS micelles around the SWNT sidewalls, resulting in the variation of percentage of m- and s-SWNTs. At low temperatures (6–10 $^{\circ}$C), the micelle structures around the SWCNTs are bigger and the SDS molecules are closely packed, possibly making them less susceptible to be physisorbed on the gel agarose surface. This mechanism could explain why most SWNTs are eluted from the separation column after flushing with an SDS solution. Only the s-SWNTs with low enough SDS encapsulation density can physisorb on the agarose surface with limited physisorption sites, resulting in high percentage of s-SWNTs being eluted from the separation column after flushing with 1% SDC. At RT or higher the SDS micelles structure encapsulating the SWCNTs becomes smaller and loosely packed in a way that increases the affinity towards SWNTs by providing more physisorption sites. Consequently, SWNTs with low micelles density and random micelles structure, which includes some m-SWNTs, are more susceptible to being physisorbed and trapped onto the agarose surface even after successive addition of 1% SDS solution into the separation column. Only m-SWNTs with large surfactant micelles density will be eluted, the rest being trapped in the gel column, yielding higher m-SWNTs purity in the bottom band. Upon adding the 1% SDC solution, the remaining trapped SWNTs in the top band are eluted. Thus, the collected top band will contain higher amount of m-SWNTs, therefore yielding lower s-SWNTs purity than its low temperature counterpart.

## CNT-based transistors

CNT-FETs are then fabricated to assess the electrical characteristics of the sorted s-SWNTs and to demonstrate its direct application as electronic devices. Each device consists of a SWNTs network contacted by an Au metal source and drain electrodes with channel width of ~500 nm on heavily doped *p*-type Si wafer with 250 nm thick thermally grown $SiO_2$ dielectric. We prepared devices based on the s-SWNT fraction from each of the four SWNT samples after MS-separation at 6 $^o$C, *i.e.* DIPS-CNT (Dev-1), AD-CNT (Dev-2), HiPco (Dev-3) and CoMoCAT (Dev-4). For comparison, we also fabricated a device based on a DIPS-CNT sample (Dev-5) processed in the same way as the ones used for Dev-1 but without filtration (enrichment). Figure 13a shows a representative scanning electron micrograph (SEM) image of an array of the devices from a single SWNT source (Dev-1) whereby the measurements were carried out on one of the electrode pairs for each SWNT sources. The densities of the SWCNT networks of all devices were controlled to be as identical as possible at approximately ~5 tubes per $\mu m^2$. The zero gate bias I-V characteristic of the five devices are shown in Figure 13b. Both Dev-1 and Dev-2 show non-linear I-V curves with maximum conductance~$2.7 \times 10^{-8}$ S. For Dev2, which consist of AD-CNT channel, the I-V curve is asymmetrical possibly due to the rectifying effect of an unbalanced Schottky barrier on one of the electrode-tube interface. The IV curves for Dev-3 and Dev-4 are also non-linear, but they exhibit lower conductance~$6.2 \times 10^{-10}$ S and ~$2.1 \times 10^{-9}$ S, respectively, with respect to Dev-1 and Dev-2. The non-linear I-V curve is expected for a typical s-SWNTs network [2]. In contrary, Dev-5 has an Ohmic linear I-V curve with higher maximum conductance (~$2.2 \times 10^{-7}$ S) with respect to the other devices, which reflects the characteristics of an un-sorted SWNTs channel where carrier transport is markedly affected by the presence of m-SWNTs.

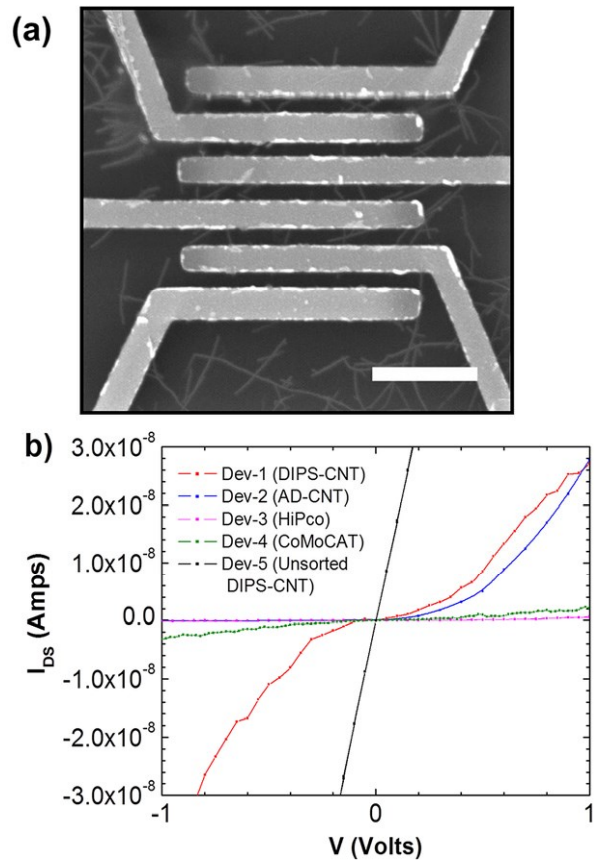

**Figure 13:** (a) SEM of representative of an array of CNT-FETs. The scale bar is 2 μm. (b) I-V curves at zero gate bias for devices based on s-SWNTs enriched channels: DIPS-CNT (Dev-1, red), AD-CNT (Dev-2, blue), HiPco (Dev-3, magenta), CoMoCAT (Dev-4, green); and un-sorted DIPS-CNT channel (Dev-5, black).

The 3-terminal field effect output and transfer characteristics are shown in Figures 14 and 15, respectively. The output characteristic of Dev-1 in Figure 14b indicates how the gate-source voltage bias, $V_{GS}$, modulates the drain-source current, $I_{DS}$. An increase of $V_{GS}$ in the negative range corresponds to an increase of $I_{DS}$ (ON state). A gradual decrease of $I_{DS}$, towards values approaching zero (OFF state), is seen when $V_{GS}$ increases in the positive range, which indicates p-type FET behavior [108]. The output characteristics of Dev-2, Dev-3 and Dev-4 are very similar to Dev-1. The un-sorted SWNTs channel of Dev-5 by comparison produces very slight gate bias modulation and higher $I_{DS}$ with respect to that of Dev-1 (Figure 14b). This is another indication that the carrier transport in the channel is dominated by m-SWNTs, which cannot be fully depleted, thus the devices cannot be turned off. The analysis of the output characteristics, and the comparison with that achieved with Dev-5, further confirms the spectroscopic analysis, indicating that the sorted SWNT network in Dev-1, Dev-2, Dev-3, Dev-4 containing mainly s-SWNTs.

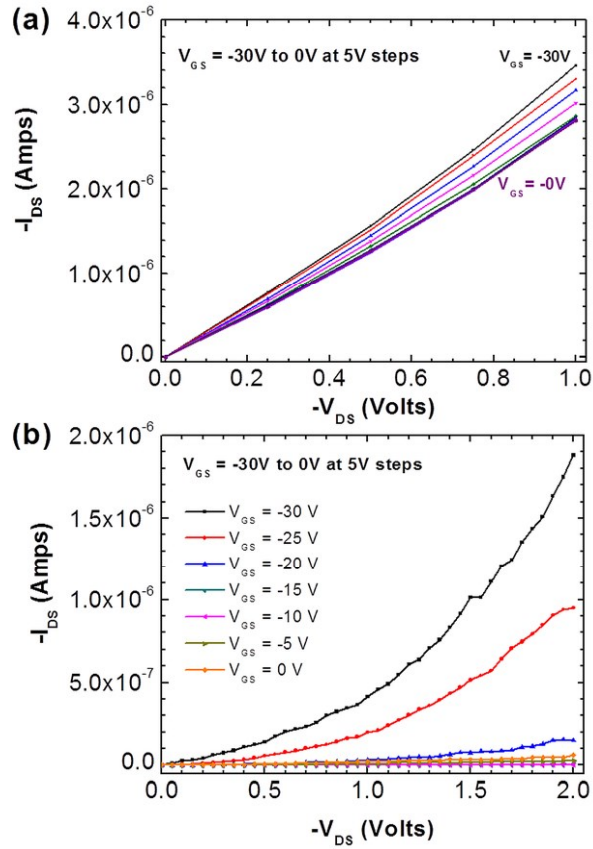

**Figure 14:** Output characteristics for (a) devices with un-sorted DIPS-CNTs, and (b) device with sorted DIPS-CNTs.

The transfer characteristics of the devices in Figure 15 demonstrate proper current modulation and switching for all four SWCNT sources. On the other hand, the transfer curve for Dev-5 with unsorted SWCNTs shows virtually no switching. The transfer curve for Dev-1 shows ambipolar behavior, which is attributed to the intrinsic property of DIPS-CNT. The low turn-on voltage observed here for Dev-1 of ~-10 V is due to the large average tube diameter ($d_t$) of DIPSCNT and hence smaller bandgaps compared to the other SWCNT sources.

The Dev-1 and Dev-2 have $I_{ON}/I_{OFF}$ ~$10^5$ and ~$10^6$, respectively. Dev-3 and Dev-4 both have $I_{ON}/I_{OFF}$ ~$10^4$. Although $I_{ON}/I_{OFF}$ ratios up to $10^7$ have been reported for CNT-FETs [109], they have thin molecular high-$k$ dielectric as gate oxide [109]. CNT-FETs fabricated from sorted SWNTs via DGU [37] and gel filtration [103], [110], [111] have shown lower $I_{ON}/I_{OFF}$ ~$10^3$-$10^4$. The very high $I_{ON}/I_{OFF}$ of Dev-2 is predominantly due to the low off-current ~$10^{-13}$ A, compared to that of Dev-1, ~$10^{-11}$ A. Dev-3 and Dev-4 also have low off-current of ~$10^{-13}$ A, but the on-currents are lower than in Dev-1 and Dev-2. This results in a lower $I_{ON}/I_{OFF}$. For Dev-2, the average off-current is ~$10^{-12}$ A. This can be ascribed to the fact that the carrier transport in the device channel is dominated by s-SWNTs and can therefore be modulated by the field generated by gate bias. The $I_{ON}/I_{OFF}$ of Dev-5 is less than 2, with high off-current of ~4 $\mu$A, which is ~7 orders of magnitude higher than Dev-2. This is related to the presence of m-SWNTs, which dominate the carrier transport. The transfer characteristic of Dev-1 shows asymmetric ambipolar behavior that is *p*-type dominant, which can be attributed to the intrinsic properties of the DIPS-SWNTs source. SWNTs synthesized via the DIPS method has been shown to generally demonstrate ambipolar behavior when exploited as the active channel of a transistor [112], whilst SWNTs in general exhibit *p*-type behavior in air [5].

There are intrinsic physical limitations to the maximum $I_{ON}/I_{OFF}$ ratio obtainable for a given supply voltage [22]. At best, $I_{DS}$ increases exponentially from $I_{OFF}$ to $I_{ON}$, with a rate described by the subthreshold swing (SS), expressed in mV dec$^{-1}$ *(i.e.,* mV of incremental gate voltage required to change the drain current by one decade) [22], [108].

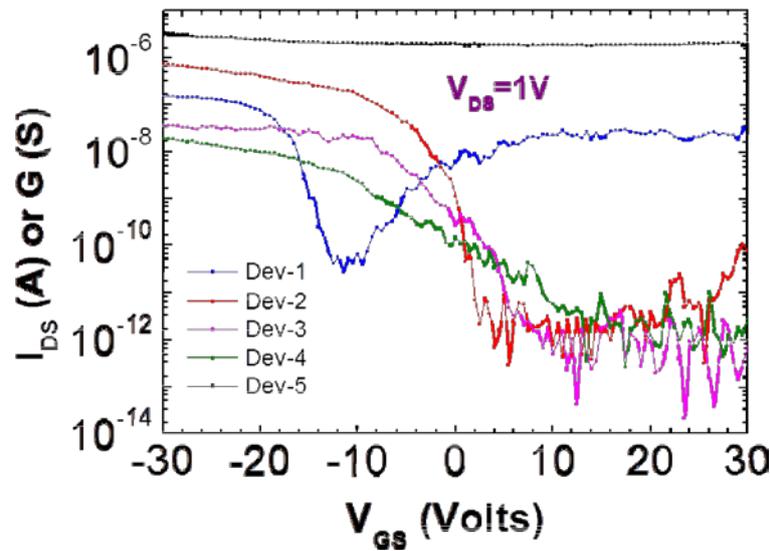

**Figure 15:** Transfer characteristics for device with sorted DIPS-CNTs channel (Dev-1), sorted AD-CNTs channel (Dev-2), sorted HiPco channel (Dev-3), sorted CoMoCAT (Dev-4) and unsorted DIPS-CNTs channel (Dev-5). The source-drain voltage, $V_{DS}$, is 1 V for all curves.

Therefore, a low *SS* is desirable for efficient device switching [108]. From Figure 15, *SS* for Dev1 to Dev-4 is estimated to be 700, 1100, 1570 and 7700 mV/dec, respectively. These values are higher to the best *SS* reported for CNT-FETs devices with similar structures (~200 mV/dec) [113]. Our higher *SS* can be attributed to the thick 250nm $SiO_2$, and lower concentration of SWNTs in the channel network [114], as well as charge in the $SiO_2$ and surface traps. Another indicator of CNT-FET switching performance is the transconductance ($g_m$), defined as $g_m = \Delta I_{OUT}/\Delta V_{IN}$, where $I_{OUT}$ is the current variation at the output and $\Delta V_{IN}$ the voltage variation at the input [108]. Dev-1 and Dev-2 have $g_m \sim 2.45 \times 10^{-8}$ S for both devices. Dev-3 and Dev-4 exhibit much lower $g_m \sim 0.3 \times 10^{-8}$ S and $\sim 0.1 \times 10^{-8}$ S, respectively. As for the case of *SS*, this low $g_m$ value can be attributed to the thick dielectric layer, which introduces high parasitic capacitances and charge trapping between channel and gate. Thus, it is not attributable to the presence of highly enriched s-SWNTs but only to device structure. However, the variation of *SS* and $g_m$ between all four devices shown here can be attributed to the respective

SWNT sources used that have varying average tube diameter, and hence band gap. It has been shown that tube diameter can influence the device performance [115], [116], [117].

The mobility ($\mu$) of our CNT-FETs can also be determined from the 3-terminal electrical characteristic measurements. Here, we can estimate the "device mobility" or "field effect mobility", $\mu_{FE}$, which takes into account the scattering effects of the SWNT channel and the metal-SWNT contacts, rather than the intrinsic mobility of SWNT alone [118], [119]. A single device is measured between a pair of source-drain electrodes with channel length of ~500 nm. The active channel of the devices consists of either a single or a few SWNTs and therefore assumed to be 1D system to simplify the calculation. The $\mu_{FE}$ can be calculated directly by using the formula:

$$\mu_{FE} = \frac{L}{C_G}\frac{\partial G}{\partial V_{GS}} \quad (1)$$

where $L$ is the SWNT channel length, $C_G$ is the gate capacitance and $G$ is the device conductance. The $C_G$ can be calculated using:

$$C_G = L\left[C_Q^{-1} + \frac{\ln(2\frac{t_{ox}}{r})}{2\pi\varepsilon_0\varepsilon_r}\right]^{-1} \quad (2)$$

where $C_Q^{-1}$ is the quantum conductance (0.4 nF.m$^{-1}$) [120], $t_{ox}$ is the oxide thickness (~250 nm) and, $\varepsilon_0$ and $\varepsilon_r$ are vacuum permittivity and SiO$_2$ relative permittivity, respectively. Parameter $r$ is the SWNT diameter, which is approximated by the average diameter of the respective SWNT sources. The maximum calculated $\mu_{FE}$ of ~49 cm$^2$V$^{-1}$s$^{-1}$ was recorded for Dev-1 (DIPS-CNT), which is the highest demonstrated here. The maximum $\mu_{FE}$ for device Dev-2 (AD-CNT), Dev-3 (HiPco) and Dev-4 (CoMoCAT) are 5.2 cm$^2$V$^{-1}$s$^{-1}$, 4.2 cm$^2$V$^{-1}$s$^{-1}$ and 2.3 cm$^2$V$^{-1}$s$^{-1}$, respectively. The values demonstrated here is typical for devices from solution-suspended SWNTs, which are generally inferior to as grown CVD SWNT devices [119]. The performance discrepency can be attributed to the high SWNT-electrode contact resistance and strong scattering in the channel region from structural defects [121] likely induced by the sonication process.

## CONCLUSIONS

We demonstrate the effect of process temperature towards the type sorting of m- and s-SWNTs via column chromatography, exploiting agarose gel beads. The sorting procedure is of general validity, being almost independent on nanotubes source. The percentage of s- and m-SWNTs can be controlled by varying the process temperature. The enrichment of s-SWNTs is more effective at low temperatures (6-10 $^o$C), while m-SWNTs are sorted more effectively within the temperature range 30-40 $^o$C. The best performances are achieved with HiPco:~ 95% s-SWNTs at 6 $^o$C. The temperature dependence of the sorting process helps understanding the sorting mechanism. It is suggested that the separation process temperature influences the SDS micelles' structure size and distribution density, consequently affecting the separation purity. The high purity achieved by our separation protocol allowed us to fabricate CNT-FETs, using sorted s-SWNTs, reaching $I_{ON}/I_{OFF}$ ~$10^6$, a further indication that the channels consist predominantly of s-SWNTs.

## ACKNOWLEDGMENTS

We thank G. Fiori for useful discussion and M. J. Beliatis for the graphics in figure 12. We acknowledge funding from a Royal Society Wolfson Research Merit Award, the European Research Council Grants NANOPOTS, Hetero2D, EU grants GENIUS, and Graphene Flagship (contract no.604391), EPSRC grants EP/K01711X/1, EP/K017144/1, EP/L016087/1, Newton International Fellowship. I.Y. is grateful to the Government of Malaysia and Universiti Kebangsaan Malaysia (GGPM-2013-073) for funding the PhD studentship and S.R.P.S acknowledges the support received from EPSRC (EP/F048068/1) in support for the work conducted.

## REFERENCES

[1] Dresselhaus MS, Dresselhaus G, Avouris P. Carbon Nanotubes: Synthesis, Structure Properties and Applications. Topics in Applied Physics. Berlin: Springer-Verlag, 2001.
[2] Zhou YX, Gaur A, Hur SH, Kocabas C, Meitl MA, Shim M, et al. p-channel, n-channel thin film transistors and p-n diodes based on single wall carbon nanotube networks. Nano Lett 2004;4.
[3] Gruner G, Carbon nanotube transistors for biosensing applications. Anal Bioanal Chem 2006; 384: 322-335.
[4] Wong H. Beyond the conventional transistor. Solid State Electron. 2005; 49: 755-762.
[5] Avouris P. Carbon nanotube electronics. Chem Phys 2002; 281: 429-445.
[6] Appenzeller J. Carbon nanotubes for high-performance electronics - Progress and prospect. P. IEEE. 2008; 96: 201-211.
[7] Martel R. Sorting carbon nanotubes for electronics. ACS Nano 2008; 2: 2195-2199.
[8] Ding L, Tselev A, Wang JY, Yuan DN, Chu HB, McNicholas TP, et al. Selective growth of well-aligned semiconducting single-walled carbon nanotubes. Nano Lett 2009; 9: 800-805.
[9] Yu B, Liu C, Hou PX, Tian Y, Li S, Liu B, et al. Bulk synthesis of large diameter semiconducting single-walled carbon nanotubes by oxygen-assisted floating catalyst chemical vapor deposition. J Am Chem Soc 2011; 133: 5232-5235.
[10] Che Y, Wang C, Liu J, Liu B, Lin X, Parker J, et al. Selective synthesis and device applications of semiconducting single-walled carbon nanotubes using isopropyl alcohol as feedstock. ACS Nano 2012; 6: 7454-7462.
[11] Song W, Jeon C, Kim YS, Kwon YT, Jung DS, Jang SW, et al. Synthesis of bandgap controlled semiconducting single-walled carbon nanotubes. ACS Nano 2010; 4: 1012-1018.
[12] Chen JS, Stolojan V, Silva SRP. Towards type-selective carbon nanotube growth at low substrate temperature via photo-thermal chemical vapour deposition. Carbon 2015; 84: 409418.
[13] Li J, Liu K, Liang S, Zhou W, Pierce M, Wang F, et al. Growth of high-density-aligned and semiconducting-enriched single-walled carbon nanotubes: decoupling the conflict between density and selectivity. ACS Nano 2014; 8: 554-562.
[14] Qin X, Peng F, Yang F, He X, Huang H, Luo D, et al. Growth of semiconducting single walled carbon nanotubes by using ceria as catalyst supports. Nano Lett 2014; 14: 512-517.
[15] Wang H, Wang B, Quek XY, Wei L, Zhao J, Li LJ, et al. Selective synthesis of (9,8) single walled carbon nanotubes on cobalt incorporated tud-1 catalyst. J Am Chem Soc 2010; 132: 16747-16749.
[16] Liu J, Wang C, Tu X, Liu B, Chen l, Zheng M, et al. Chirality-controlled synthesis of single-wall carbon nanotubes using vapour-phase epitaxy. Nat Commun 2012; 3: 1199.

[17] Novoselov KS, Geim AK. The rise of graphene, Nat Mater 2007; 6: 183-191.
[18] Bonaccorso F, Sun Z, Hasan T, Ferrari AC. Graphene photonics and optoelectronics. Nat Photonics 2010; 4: 611-622.
[19] Ferrari AC, Bonaccorso F, Falko V, Novoselov KS, Roche S, Boggild P, et al. Science and technology roadmap for graphene, related two-dimensional crystals, and hybrid systems. Nanoscale 2014; DOI: 10.1039/C4NR01600A.
[20] Wang QH, Kalantar-zadeh K, Kis A, Coleman JN, Strano MS. Electronics and optoelectronics of two-dimensional transition metal dichalcogenides. Nat Nanotechnol 2012; 7: 699-712.
[21] Bonaccorso F, Lombardo A, Hasan T, Sun Z, Colombo L, Ferrari AC. Production and processing of graphene and 2d crystals. Mater Today 2012; 15: 564-589.
[22] Fiori G, Bonaccorso F, Iannaccone G, Palacios T, Neumaier D, Seabaugh A, et al. Electronics based on two-dimensional materials. Nat Nanotechnol 2014; 9: 768-779.
[23] Krupke R, Hennrich F, von Lohneysen H, Kappes MM. Separation of metallic from semiconducting single-walled carbon nanotubes. Science 2003; 301: 344-347.
[24] Zheng M, Jagota A, Strano MS, Santos AP, Barone P, Chou SG, et al. Structure-based carbon nanotube sorting by sequence-dependent DNA assembly. Science 2003; 302: 15451548.
[25] Collins P, Arnold M, Avouris P. Engineering carbon nanotubes and nanotube circuits using electrical breakdown. Science 2001; 292: 706-709.
[26] Arnold MS, Green, A.A.; Hulvat, J.F.; Stupp, S.I.; Hersam, M.C. Sorting carbon nanotubes by electronic structure using density differentiation. Nat. Nanotechnol. 2006, 1, 60-65.
[27] Tanaka T, Jin HH, Miyata Y, Kataura H. High-yield separation of metallic and semiconducting single-wall carbon nanotubes by agarose gel electrophoresis. Appl Phys Express. 2008; 1.
[28] Hasan T, Sun Z, Wang F, Bonaccorso F, Tan PH, Rozhin AG, Ferrari AC. Nanotubepolymer composites for ultrafast photonics. Adv Mater 2009; 21: 3874-3899.
[29] Rozhin AG, Scardaci V, Wang F, Hennrich F, White IH, Milne WI, et al. Generation of ultra-fast laser pulses using nanotube mode-lockers. Phys Status Solidi B 2006; 243: 35513555.
[30] Wang F, Rozhin AG, Scardaci V, Sun Z, Hennrich F, White IH, et al. Wideband-tuneable, nanotube mode-locked, fibre laser. Nat Nanotechnol 2008; 3: 738-742.
[31] Bonaccorso F, Hasan T, Tan PH, Sciascia C, Privitera G, Di Marco G, et al. Density gradient ultracentrifugation of nanotubes: interplay of bundling and surfactants encapsulation. J. Phys Chem C 2010; 114: 17267-17285.
[32] Gosh S, Bachilo SM, Weisman RB. Advanced sorting of single-walled carbon nanotubes by nonlinear density gradient ultracentrifugation. Nat Nanotechnol 2010; 5: 443-450.
[33] Ju SY, Utz M, Papadimitrakopoulos F. Enrichment mechanism of semiconducting singlewalled carbon nanotubes by surfactant amines. J Am Chem Soc 2009; 131: 6775-6784.
[34] Hersam MC. Progress towards monodisperse single-walled carbon nanotubes. Nat Nanotechnol 2008; 3: 387-394.
[35] Liu J, Hersam MC. Recent developments in carbon nanotube sorting and selective growth. MRS Bull 2010; 35: 315-321.
[36] Fagan JA, Becker ML, Chun J, Nie P, Bauer BJ, Simpson JR, et al. Centrifugal length separation of carbon nanotubes. Langmuir 2008; 24: 13880-13889.
[37] Arnold MS, Stupp SI, Hersam MC. Enrichment of single-walled carbon nanotubes by diameter in density gradients. Nano Lett 2005; 5: 713-718.
[38] Green AA, Duch MC, Hersam MC. Isolation of single-walled carbon nanotube enantiomers by density differentiation. Nano Res 2009; 2: 69-77.
[39] Crochet J, Clemens M, Hertel T. Quantum yield heterogeneities of aqueous single-wall carbon nanotube suspensions. J Am Chem Soc 2007; 129: 8058-+.
[40] Liu CH, Zhang HL. Chemical approaches towards single-species single-walled carbon nanotubes. Nanoscale 2010; 2.
[41] Campidelli S, Meneghetti M, Prato M. Separation of metallic and semiconducting singlewalled carbon nanotubes via covalent functionalization. Small 2007; 3.
[42] Hwang J, Nish A, Doig J, Douven S, Chen C, Chen L, et al. Polymer structure and solvent effects on the selective dispersion of single-walled carbon nanotubes. J Am Chem Soc 2008; 130: 3543-3553.

[43]  Hwang J, Kim DS, Ahn D, Hwang SW. Transport properties of a DNA-conjugated singlewall carbon nanotube field-effect transistor. Jpn J Appl Phys 2009; 48: 4.

[44]  Schug TT, Abagyan R, Blumberg B, Collins TJ, Crews D, DeFur PL, et al. Designing endocrine disruption out of the next generation of chemicals. Green Chem 2013; 15: 181-198.

[45]  Ali-Boucetta H, Nunes A, Sainz R, Herrero MA, Tian B, M. Prato M, et al. Asbestos-like pathogenicity of long carbon nanotubes alleviated by chemical functionalization. Ang Chem Int Ed 2013; 52: 2274-2278.

[46]  Tanaka T, Urabe Y, Nishide D, Kataura H. Continuous separation of metallic and semiconducting carbon nanotubes using agarose gel. Appl Phys Express 2009; 2.

[47]  Flavel BS, Moore KE, Pfohl M, Kappes MM, Hennrich F. Separation of single-walled carbon nanotubes with a gel permeation chromatography system. ACS Nano 2014; 8: 18171826.

[48]  Liu H, Nishide D, Tanaka T, Kataura H. Large-scale single-chirality separation of singlewall carbon nanotubes by simple gel chromatography. Nat Commun 2011; 2: 309.

[49]  Hirano A, Tanaka T, Kataura H. Thermodynamic determination of the metal/semiconductor separation of carbon nanotubes using hydrogels. ACS Nano 2012; 6: 10195-10205.

[50]  Ago H, Ohshima S, Uchida K, Yumura M. Gas-phase synthesis of single-wall carbon nanotubes from colloidal solution of metal nanoparticles. J Phys Chem B 2001; 105: 10453-10456.

[51]  Saito T, Xu W, Ohshima S, Ago H, Yumura M, Iijima S. Supramolecular catalysts for the gas-phase synthesis of single-walled carbon nanotubes. J Phys Chem B 2006; 110: 5849-5853.

[52]  Shi ZJ, Lian YF, Zhou XH, Gu ZN, Zhang YG, Iijima S, et al. Mass-production of singlewall carbon nanotubes by arc discharge method. Carbon 1999; 37.

[53]  Kitiyanan B, Alvarez WE, Harwell JH, Resasco DE. Controlled production of single-wall carbon nanotubes by catalytic decomposition of CO on bimetallic Co-Mo catalysts. Chem PhysLett 2000; 317: 497-503.

[54]  Resasco DE, Alvarez WE, Pompeo F, Balzano L, Herrera JE, Kitiyanan B, et al. A scalable process for production of single-walled carbon nanotubes (SWNTs) by catalytic disproportionation of CO on a solid catalyst. J Nanopart Res 2002; 4: 131-136.

[55]  Bronikowski MJ, Willis PA, Colbert DT, Smith KA, Smalley RE. Gas-phase production of carbon single-walled nanotubes from carbon monoxide via the HiPco process: A parametric study. J Vac Sci Technol A 2001; 19: 1800-1805.

[56]  O'Connell M, Bachilo S, Huffman C, Moore V, Strano M, Haroz E, et al. Band gap fluorescence from individual single-walled carbon nanotubes. Science 2002; 297: 593-596.

[57]  Tan PH, Rozhin AG, Hasan T, Hu P, Scardaci V, Milne WI, et al. Photoluminescence spectroscopy of carbon nanotube bundles: Evidence for exciton energy transfer. Phys Rev Lett 2007; 99.

[58]  Tan PH, Hasan T, Bonaccorso F, Scardaci V, Rozhin AG, Milne WI, et al. Optical properties of nanotube bundles by photoluminescence excitation and absorption spectroscopy. Physica E 2008; 40: 2352-2359.

[59]  Hagen A, Hertel T. Quantitative analysis of optical spectra from individual single-wall carbon nanotubes. Nano Lett 2003; 3: 383-388.

[60]  Hasan T, Scardaci V, Tan P, Rozhin AG, Milne WI, Ferrari AC. Stabilization and "Debundling" of single-wall carbon nanotube dispersions in N-Methyl-2-pyrrolidone (NMP) by polyvinylpyrrolidone (PVP). J Phys Chem C 2007; 111: 12594-12602.

[61]  Workman JJ. Applied Spectroscopy: A Compact Reference for Practitioners. San Diego: Academic Press; 1998.

[62]  Krupke R, Hennrich F, Hampe O, Kappes MM. Near-infrared absorbance of single-walled carbon nanotubes dispersed in dimethylformamide. J Phys Chem B 2003; 107: 5667-5669.

[63]  Landi BJ, Ruf HJ, Worman JJ, Raffaelle RP. Effects of alkyl amide solvents on the dispersion of single-wall carbon nanotubes. J Phys Chem B 2004; 108: 17089-17095.

[64]  Weisman RB, Bachilo SM. Dependence of optical transition energies on structure for single-walled carbon nanotubes in aqueous suspension: an empirical Kataura plot. Nano Lett 2003; 3: 1235-1238.

[65]  Kataura H, Kumazawa Y, Maniwa Y, Umezu I, Suzuki S, Ohtsuka Y, et al. Optical properties of single-wall carbon nanotubes. Synthetic Met 1999; 103: 2555-2558.

[66] Bachilo S, Strano M, Kittrell C, Hauge R, Smalley R, Weisman R. Structure-assigned optical spectra of single-walled carbon nanotubes. Science 2002; 298: 2361-2366.
[67] Maeda Y, Kimura S, Hirashima Y, Kanda M, Lian YF, Wakahara T, et al. Dispersion of single-walled carbon nanotube bundles in nonaqueous solution. J Phys Chem B 2004; 108: 18395-18397.
[68] McDonald TJ, Blackburn JL, Metzger WK, Rumbles G, Heben MJ. Chiral-selective protection of single-walled carbon nanotube photoluminescence by surfactant selection. J. Phys Chem C 2007; 111: 17894-17900.
[69] Lefebvre J, Finnie P. Polarized photoluminescence excitation spectroscopy of singlewalled carbon nanotubes. Phys Rev Lett 2007; 98.
[70] Kato T, Hatakeyama R. Exciton energy transfer-assisted photoluminescence brightening from freestanding single-walled carbon nanotube bundles. J Am Chem Soc 2008; 130: 81018107.
[71] Torrens ON, Zheng M, Kikkawa JM. Energy of K-Momentum Dark Excitons in Carbon Nanotubes by Optical Spectroscopy. Phys Rev Lett 2008; 101.
[72] Schoeppler F, Mann C, Hain TC, Neubauer FM, Privitera G, Bonaccorso F, et al. Molar Extinction Coefficient of Single-Wall Carbon Nanotubes. J Phys Chem C 2011; 115: 1468214686.
[73] Reich S, Thomsen C, Robertson J. Exciton resonances quench the photoluminescence of zigzag carbon nanotubes. Phys Rev Lett 2005; 95.
[74] Tsyboulski DA, Rocha JDR, Bachilo SM, Cognet L, Weisman RB. Structure-dependent fluorescence efficiencies of individual single-walled cardon nanotubes. Nano Lett 2007; 7: 3080-3085.
[75] Rao AM, Richter E, Bandow S, Chase B, Eklund PC, Williams KA, et al. Diameterselective Raman scattering from vibrational modes in carbon nanotubes. Science 1997; 275: 187-191.
[76] Telg H, Maultzsch J, Reich S, Hennrich F, Thomsen C. Chirality distribution and transition energies of carbon nanotubes. Phys Rev Lett 2004; 93.
[77] Meyer JC, Paillet M, Michel T, Moreac A, Neumann A, Duesberg GS, et al. Raman modes of index-identified freestanding single-walled carbon nanotubes. Phys Rev Lett 2005; 95.
[78] Fantini C, Jorio A, Souza M, Strano MS, Dresselhaus MS, Pimenta MA. Optical transition energies for carbon nanotubes from resonant Raman spectroscopy: Environment and temperature effects. Phys Rev Lett 2004; 93.
[79] Paillet M, Michel T, Meyer JC, Popov VN, Henrard L, Roth S, et al. Raman active phonons of identified semiconducting single-walled carbon nanotubes. Phys Rev Lett 2006; 96.
[80] Jorio A, Saito R, Hafner J, Lieber C, Hunter M, McClure T, et al. Structural (n, m) determination of isolated single-wall carbon nanotubes by resonant Raman scattering. Phys Rev Lett 2001; 86: 1118-1121.
[81] Araujo PT, Doorn SK, Kilina S, Tretiak S, Einarsson E, Maruyama S, et al. Third and fourth optical transitions in semiconducting carbon nanotubes. Phys Rev Lett 2007; 98.
[82] Araujo PT, Jorio A, Dresselhaus MS, Sato K, Saito R. Diameter dependence of the dielectric constant for the excitonic transition energy of single-wall carbon nanotubes. Phys Rev Lett 2009; 103.
[83] Ferrari AC, Robertson J. Interpretation of Raman Spectra of disordered and amorphous carbon. Phys Rev B 2000; 61: 14095.
[84] Ferrari AC, Robertson J. Resonant Raman spectroscopy of disordered, amorphous and diamond-like carbon. Phys Rev B 2001; 64: 075414.
[85] Ferrari AC, Basko D. Raman spectroscopy as a versatile tool for studying the properties of graphene. Nat Nanotechnol 2013; 8: 235-246.
[86] Piscanec S, Lazzeri M, Robertson J, Ferrari AC, Mauri F. Optical phonons in carbon nanotubes: Kohn anomalies, Peierls distortions, and dynamic effects. Phys Rev B 2007; 75.
[87] Jorio A, Souza AG, Dresselhaus G, Dresselhaus MS, Swan AK, Unlu MS, et al. G-band resonant Raman study of 62 isolated single-wall carbon nanotubes. Phys Rev B 2002; 65.
[88] Lazzeri M, Piscanec S, Mauri F, Ferrari AC, Robertson J. Phonon linewidths and electronphonon coupling in graphite and nanotubes. Phys Rev B 2006; 73.
[89] Tsang JC, Freitag M, Perebeinos V, Liu J, Avouris P. Doping and phonon renormalization in carbon nanotubes. Nat Nanotechnol 2007; 2: 725-730.

[90] Das A, Pisana S, Chakraborty B, Piscanec S, Saha SK, Waghmare UV, et al. Monitoring dopants by Raman scattering in an electrochemically top-gated graphene transistor. Nat Nanotechnol 2008; 3: 210-215.

[91] Das A, Sood AK, Govindaraj A, Saitta AM, Lazzeri M, Mauri F, et al. Doping in carbon nanotubes probed by Raman and transport measurements. Phys Rev Lett 2007; 99.

[92] Shiozawa H, Pichler T, Kramberger C, Rummeli M, Batchelor D, Liu Z, et al. Screening the missing electron: nanochemistry in action. Phys Rev Lett 2009; 102.

[93] Tian Y, Jiang H, von Pfaler J, Zhu Z, Nasibulin AG, Nikitin T, et al. Analysis of the size distribution of single-walled carbon nanotubes using optical absorption spectroscopy. J Phys Chem Lett 2010; 1: 1143-1148.

[94] Piscanec S, Lazzeri M, Mauri F, Ferrari AC. Optical phonons of graphene and nanotubes. Eur Phys J-Spec Top 2007; 148: 159-170.

[95] Yanagi K, Miyata Y, Tanaka T, Fujii S, Nishide D, Kataura H. Colors of carbon nanotubes. Diam Relat Mater 2009; 18: 935-939.

[96] Moshammer K, Hennrich F, Kappes MM. Selective Suspension in Aqueous Sodium Dodecyl Sulfate According to Electronic Structure Type Allows Simple Separation of Metallic from Semiconducting Single-Walled Carbon Nanotubes. Nano Res 2009; 2: 599-606.

[97] Hirano A, Tanaka T, Urabe Y, Kataura H. pH- and solute-dependent adsorption of singlewall carbon nanotubes onto hydrogels: mechanistic insights into the metal/semiconductor separation. ACS Nano 2013; 7: 10285-10295.

[98] Blanch AJ, Quinton JS, Shapter JG. The role of sodium dodecyl sulfate concentration in the separation of carbon nanotubes using gel chromatography. Carbon 2013; 60: 471–480.

[99] Clar JG, Silvera batista CA, Youn S, Bonzongo JC, Ziegler KJ. Interactive forces between sodium dodecyl sulfate-suspended single-walled carbon nanotubes and agarose gels. J Am Chem Soc 2013; 135: 17758-17767.

[100] Silvera-Batista CA, Scott DC, McLeod SM, Ziegler KJ. A mechanistic study of the selective retention of sds-suspended single-wall carbon nanotubes on agarose gels. J Phys Chem C 2011; 115: 9361-9369.

[101] Duan WH, Wang Q, Collins F. Dispersion of carbon nanotubes with SDS surfactants: a study from a binding energy perspective. Chem Sci 2011; 2: 1407-1413.

[102] Islam M, Rojas E, Bergey D, Johnson A, Yodh A. High weight fraction surfactant solubilization of single-wall carbon nanotubes in water. Nano Lett 2003; 3: 269-273.

[103] Tanaka T, Jin H, Miyata Y, Fujii S, Suga H, Naitoh Y, et al. Simple and scalable gelbased separation of metallic and semiconducting carbon nanotubes. Nano Lett 2009; 9: 14971500.

[104] Xu Z, Yang X, Yang Z. A molecular simulation probing of structure and interaction for supramolecular sodium dodecyl sulfate/single-wall carbon nanotube assemblies. Nano Lett 2010; 10: 985-991.

[105] Tummala NR, Striolo A. SDS surfactants on carbon nanotubes: aggregate morphology. ACS Nano 2009; 3: 595-602.

[106] Wang RK, Chen WC, Campos DK, Ziegler KJ. Swelling the micelle core surrounding single-walled carbon nanotubes with water-immiscible organic solvents. J Am Chem Soc 2008; 130: 16330-16337.

[107] Hammouda B. Temperature effect on the nanostructure of SDS micelles in water. J Res Natl Inst Stand Technol 2013; 118: 151-167.

[108] Sze SM. Physics of Semiconductor Devices, 3rd ed. New York: Wiley; 2007: 6.2.4, p. 315.

[109] Weitz RT, Zschieschang U, Forment-Aliaga A, Kaelblein D, Burghard M, Kern K, et al. Highly reliable carbon nanotube transistors with patterned gates and molecular gate dielectric. Nano Lett 2009; 9: 1335-1340.

[110] Wu J, Jiao L, Antaris A, Choi CL, Xie L, Wu Y, et al. Self-Assembly of Semiconducting Single-Walled Carbon Nanotubes into Dense, Aligned Rafts. Small 2013; 9: 4142–4148.

[111] Ostermaier F, Mertig M. Phys Status Solidi B 2013; 250: 2564-2568.

[112] Yahya I, Stolojan V, Clowes S, Mustaza SM, Silva SRP. Carbon nanotube field effect transistor measurements in vacuum. Proceedings, IEEE Int. Conf. Semicond. Electron. (Melaka. Malaysia): 2010; p. 224-228.

[113] Cao Q, Kim HS, Pimparkar N, Kulkarni JP, Wang C, Shim M, et al. Medium-scale carbon nanotube thin-film integrated circuits on flexible plastic substrates. Nature 2008; 454: 495-U4.

[114] Nihey F, Hongo H, Yudasaka M, Iijima S. A top-gate carbon-nanotube field-effect transistor with a titanium-dioxide insulator. Jpn J Appl Phys 2002; 41(2).

[115] Tseng Y, Phoa K, Carlton D, Bokor J. Effect of diameter variation in a large set of carbon nanotube transistors. Nano Lett 2006; 6(7): 1364-1368.

[116] Asada Y, Nihey F, Ohmori S, Shinohara H, Saito T. Diameter-dependent performance of single-walled carbon nanotube thin-film transistors. Adv Mater 2011; 23: 4631-4635.

[117] Islam AE, Du F, Ho X, Jin SH, Dunham S, Rogers JA. Effects of variations in diameter and density on the statistics of aligned carbon-nanotube field effect transistors. J Appl Phys 2012; 111: 054511.

[118] Durkop T, Kim BM, Fuhrer MS. Properties and applications of high-mobility semiconducting nanotubes. J Phys-Condens Mat 2004; 16: R553-R580.

[119] Cao Q, Han SJ, Tulevski GS, Franklin AD, Haensch W. Evaluation of Field-Effect Mobility and Contact Resistance of Transistors That Use Solution-Processed Single-Walled Carbon Nanotubes. ACS Nano 2012; 6: 6471-6477.

[120] Rosenblatt S, Yaish Y, Park J, Gore J, Sazonova V, McEuen PL. High performance electrolyte gated carbon nanotube transistors. Nano Lett 2002; 2: 869-872.

[121] Neophytou N, Kienle D, Polizzi E, Anantram MP. Influence of defects on nanotube transistor performance. Appl Phys Lett 2006; 88(24): 242106.

Supporting Information

# Temperature Dependent Separation of Metallic and Semiconducting Carbon

*Iskandar Yahya[1]‡, Francesco Bonaccorso[2,3], Steven K. Clowes[1], Andrea C. Ferrari[2], and S. R. P. Silva[1]\**

[1] Advanced Technology Institute, University of Surrey, Guildford GU2 7XH, UK.

[2] Cambridge Graphene Centre, Cambridge University, Cambridge CB3 0FA, UK.

[3] Materials Characterization Facility, Istituto Italiano di Tecnologia, via Morego 30, 16163 Genova, Italy.

‡ Currently at the Institute of Microengineering and Nanoelectronics, Level 4, Research Complex, Universiti Kebangsaan Malaysia, 43600, Bangi, Malaysia.

\* Corresponding author: s.silva@surrey.ac.uk.

## S1. Raman spectra for DIPS-CNTs

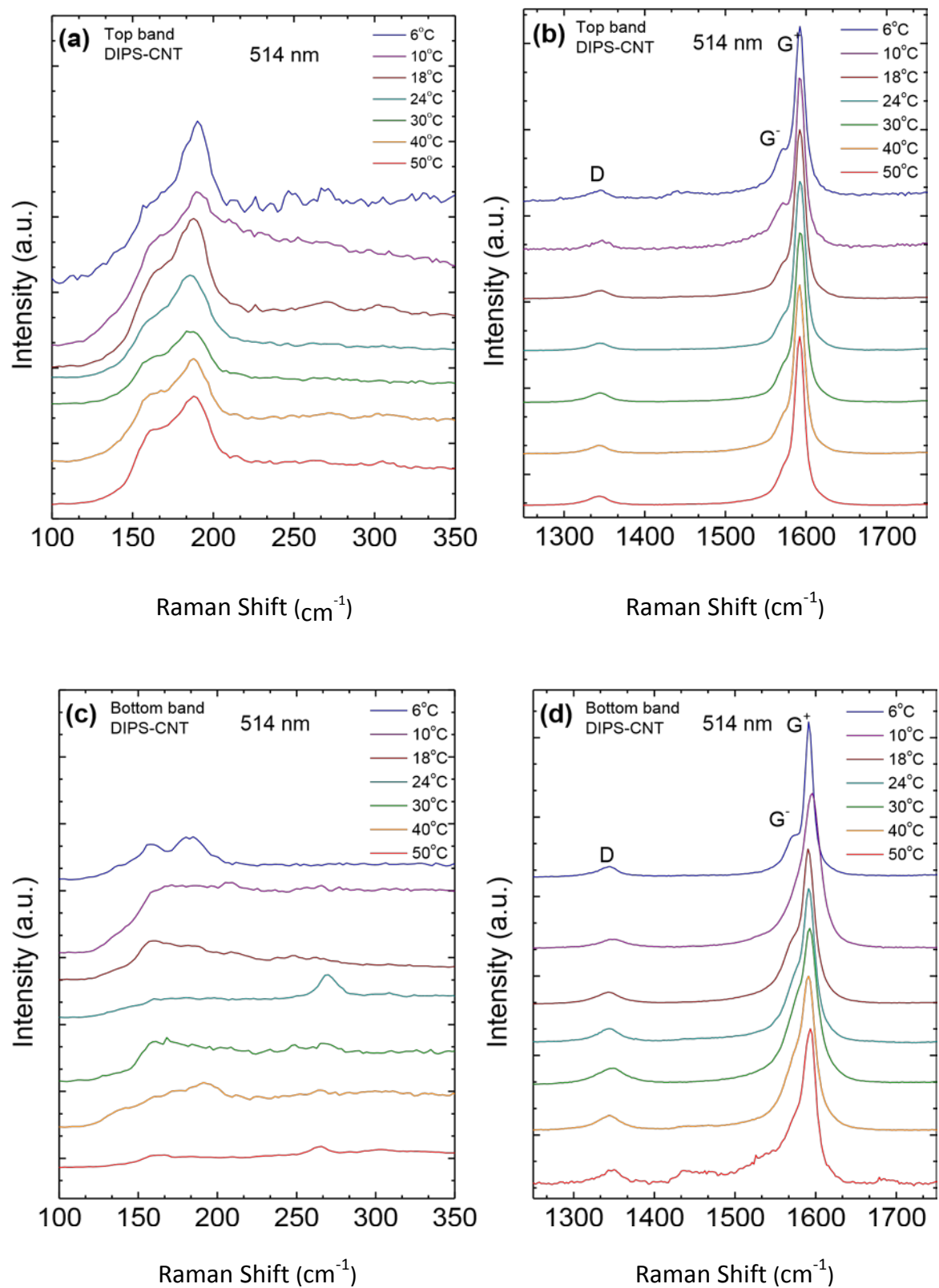

**Fig. S1.1**: Raman spectra of DIPS-CNTs separated at different temperatures; excitation wavelength 514 nm. (a) RBM and (b) D-G regions for top band; (c) RBM and (d) D-G regions for bottom band.

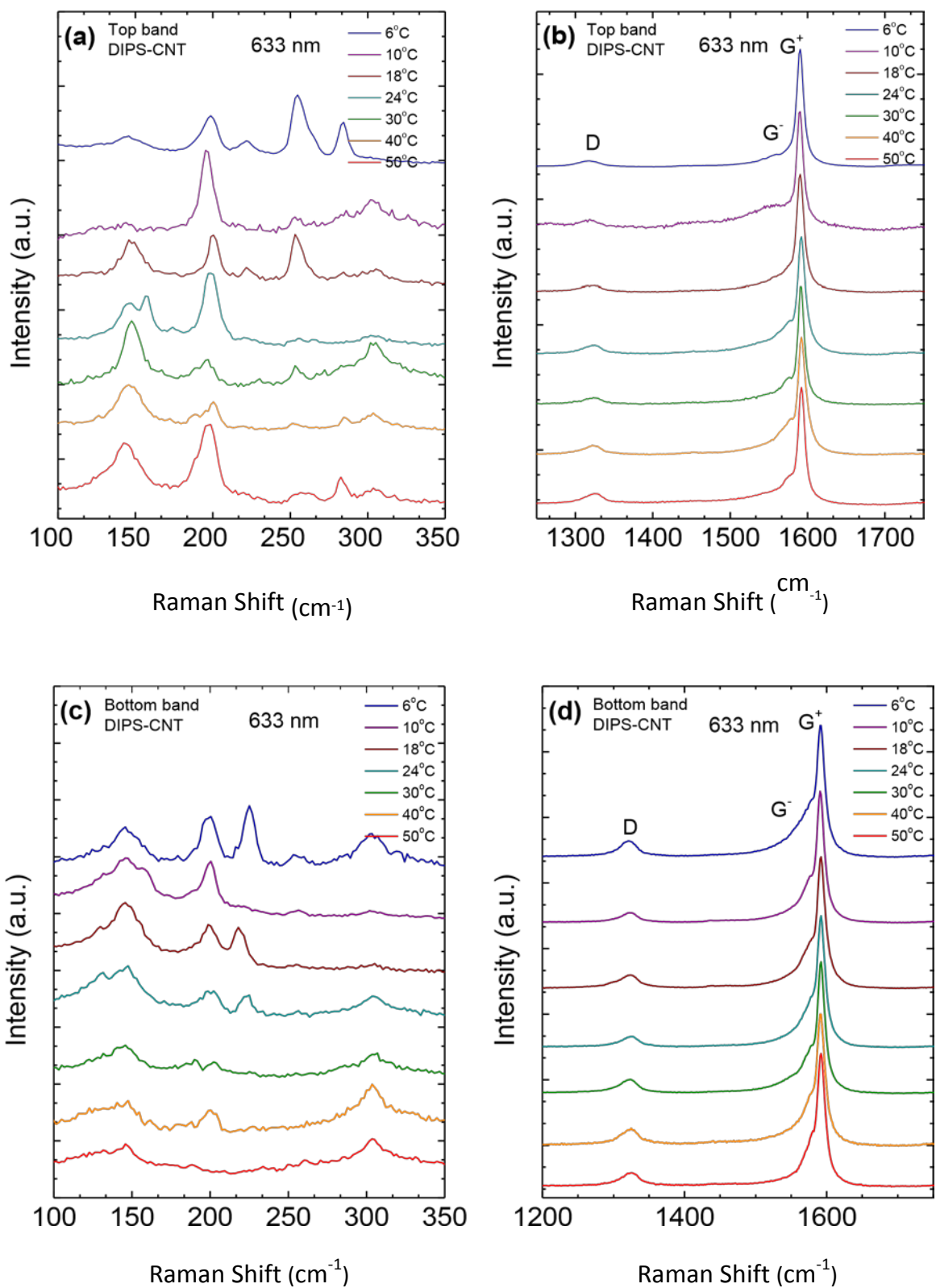

Fig. S1.2: Raman spectra of DIPS-CNTs separated at different temperatures; excitation wavelength 633 nm. (a) RBM and (b) D-G regions for top band; (c) RBM and (d) D-G regions for bottom band.

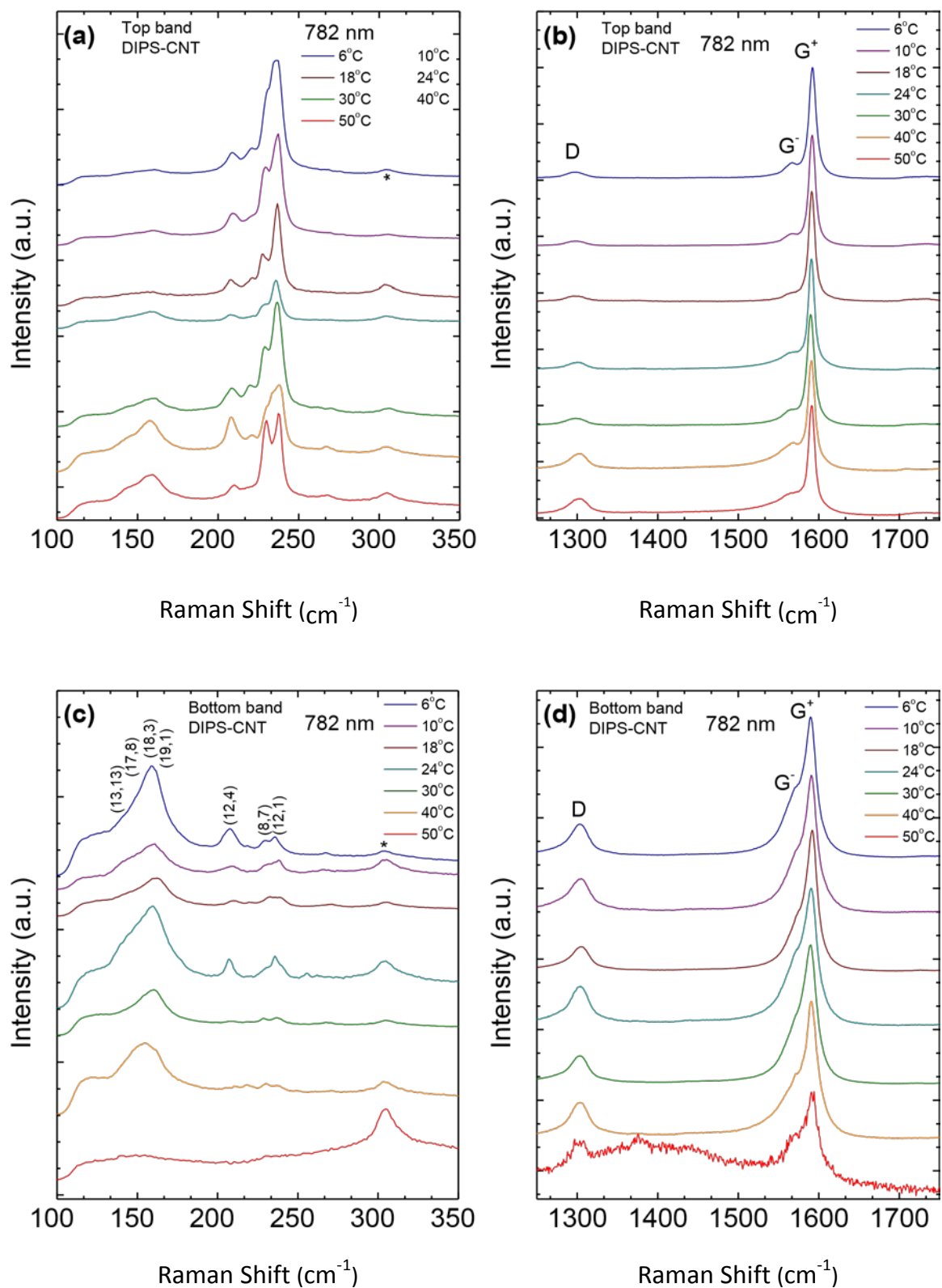

Fig. S1.3: Raman spectra of DIPS-CNTs separated at different temperatures; excitation wavelength 782 nm. (a) RBM and (b) D-G regions for top band; (c) RBM and (d) D-G regions for bottom band.

S2 Raman spectra for AD-CNTs

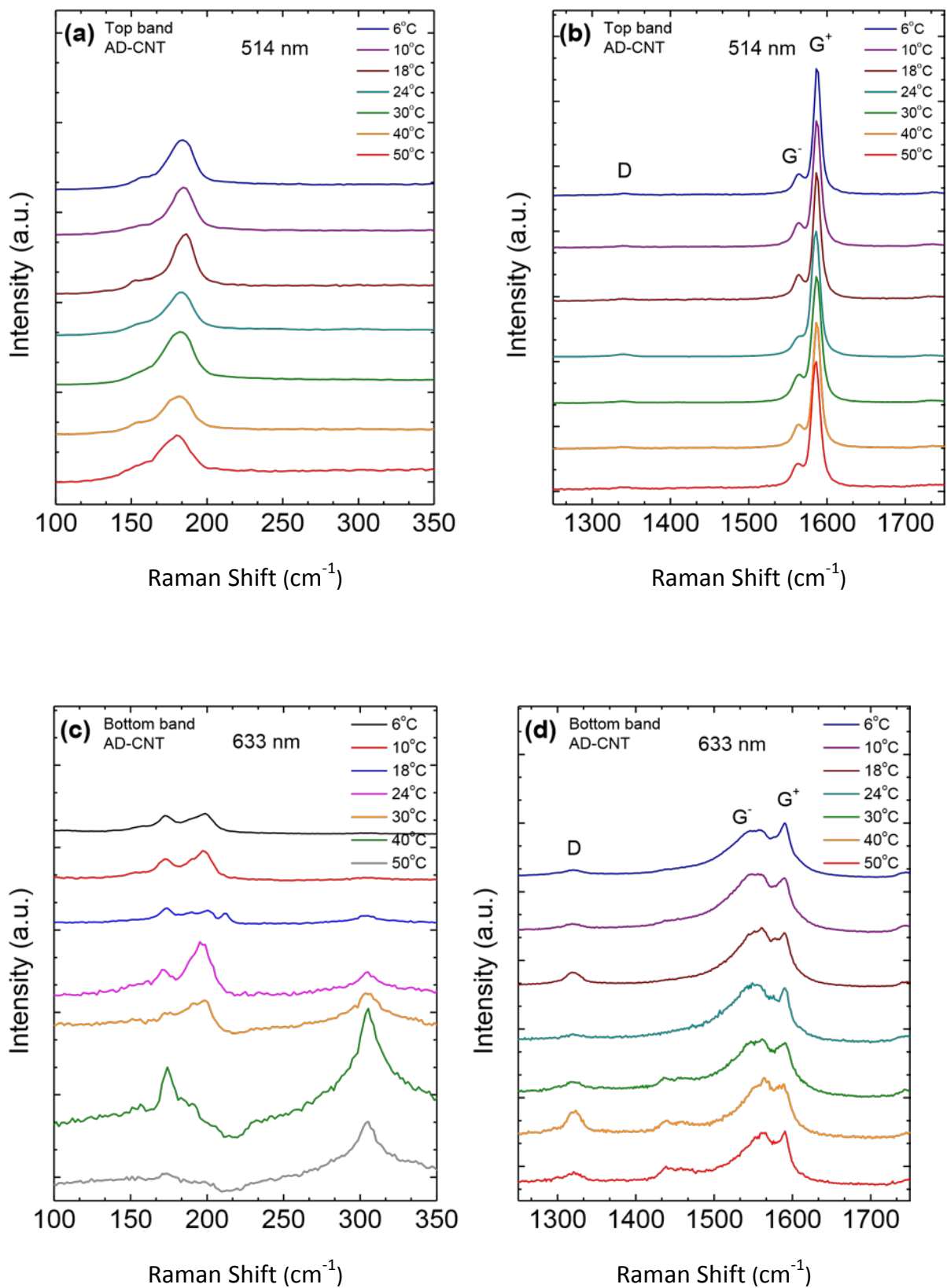

Figure S2.1: Raman spectra of AD-CNTs separated at different temperatures; excitation wavelength 514 nm. (a) RBM and (b) D-G regions for top band; (c) RBM and (d) D-G regions for bottom band.

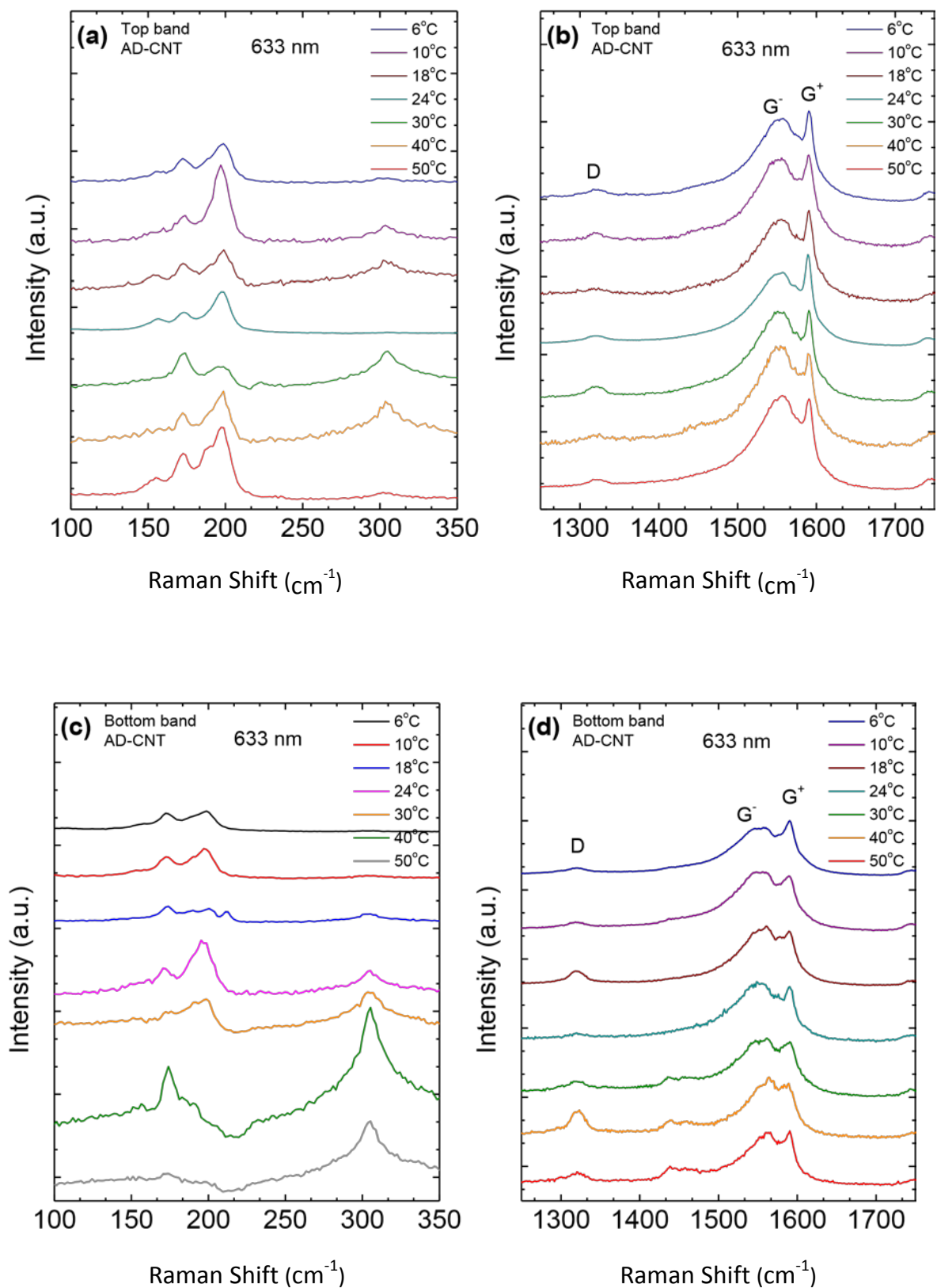

Fig. S2.2: Raman spectra of AD-CNTs separated at different temperatures; excitation wavelength 633 nm. (a) RBM and (b) D-G regions for top band; (c) RBM and (d) D-G regions for bottom band.

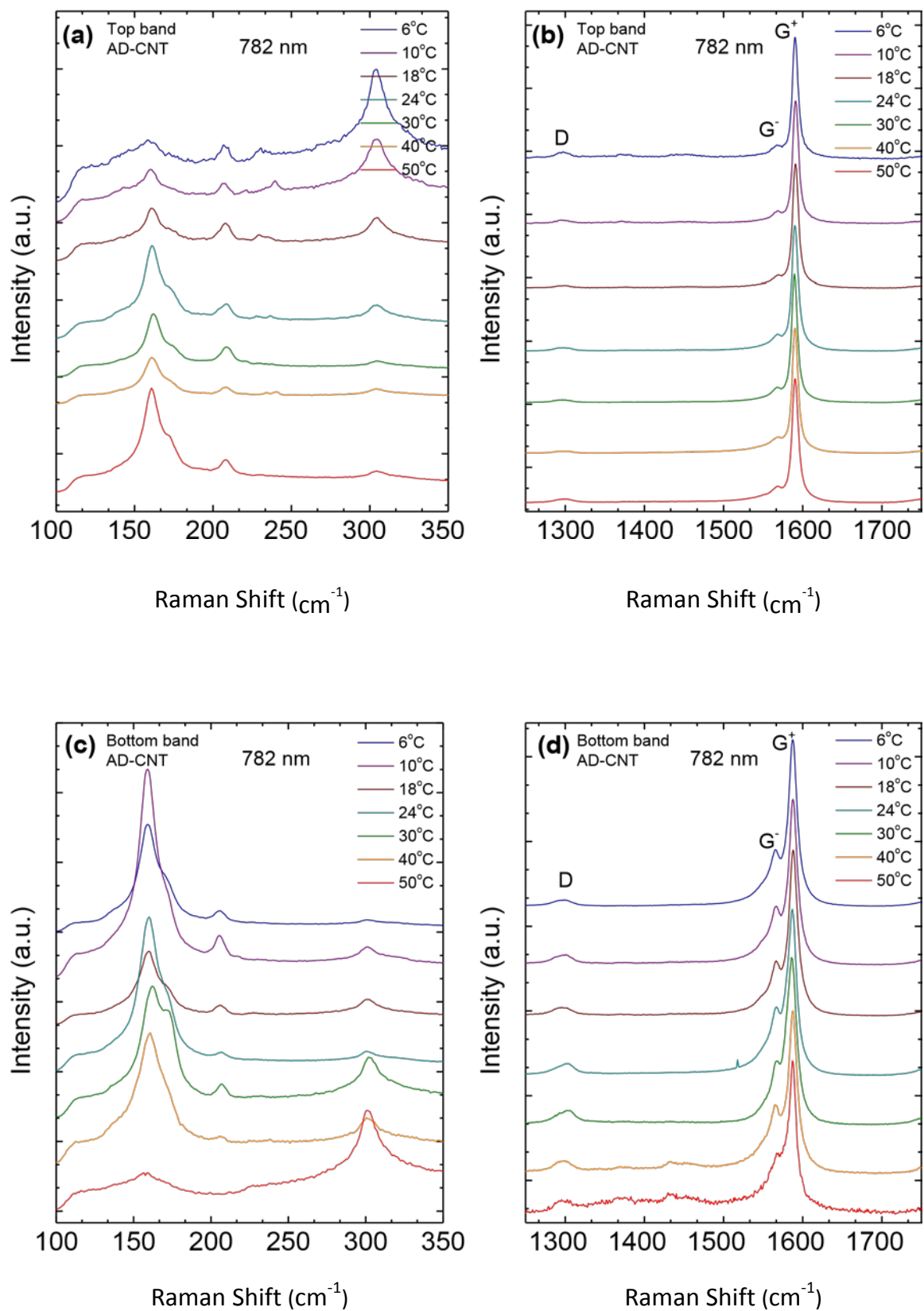

Fig. S2.3: Raman spectra of AD-CNTs separated at different temperatures; excitation wavelength 782 nm. (a) RBM and (b) D-G regions for top band; (c) RBM and (d) D-G regions for bottom band.

## S3. Raman spectra of HiPco-SWNTs

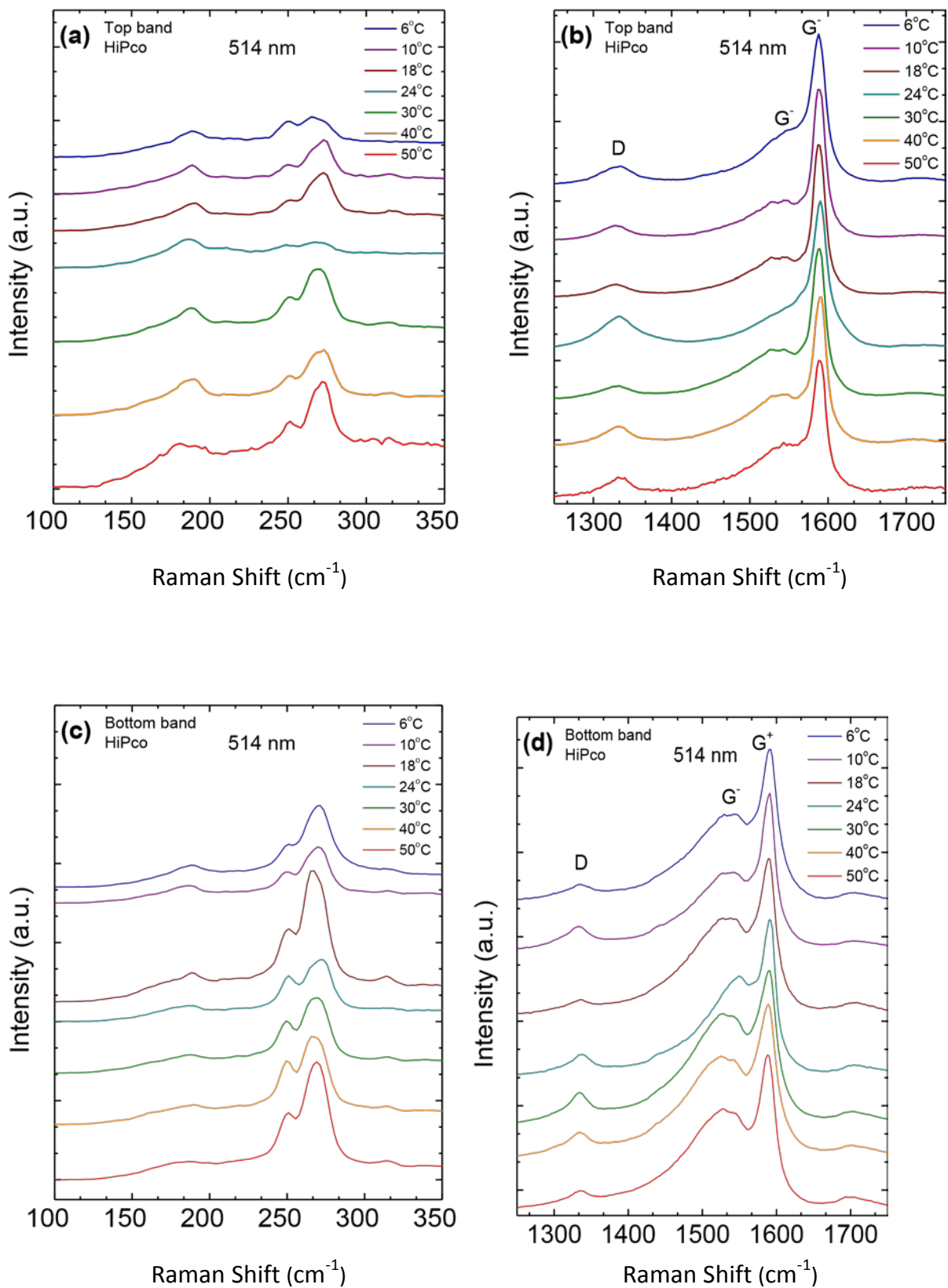

Fig. S3.1: Raman spectra of sample HiPco separated at different temperatures; excitation wavelength 514 nm. (a) RBM and (b) D-G regions for top band; (c) RBM and (d) D-G regions for bottom band.

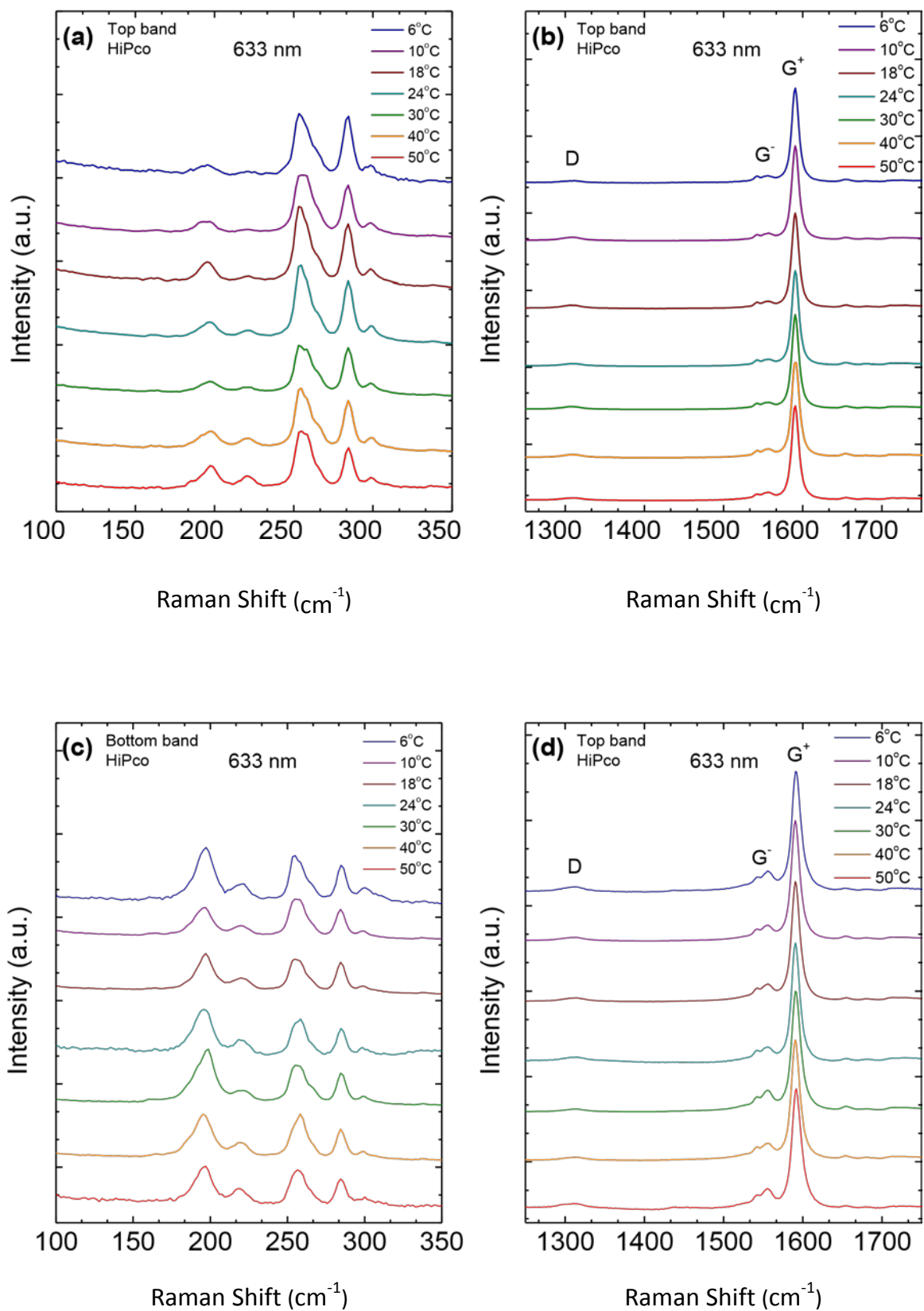

Fig. S3.2: Raman spectra of HiPco-SWNTs separated at different temperatures; excitation wavelength 633 nm. (a) RBM and (b) D-G regions for top band; (c) RBM and (d) D-G regions for bottom band.

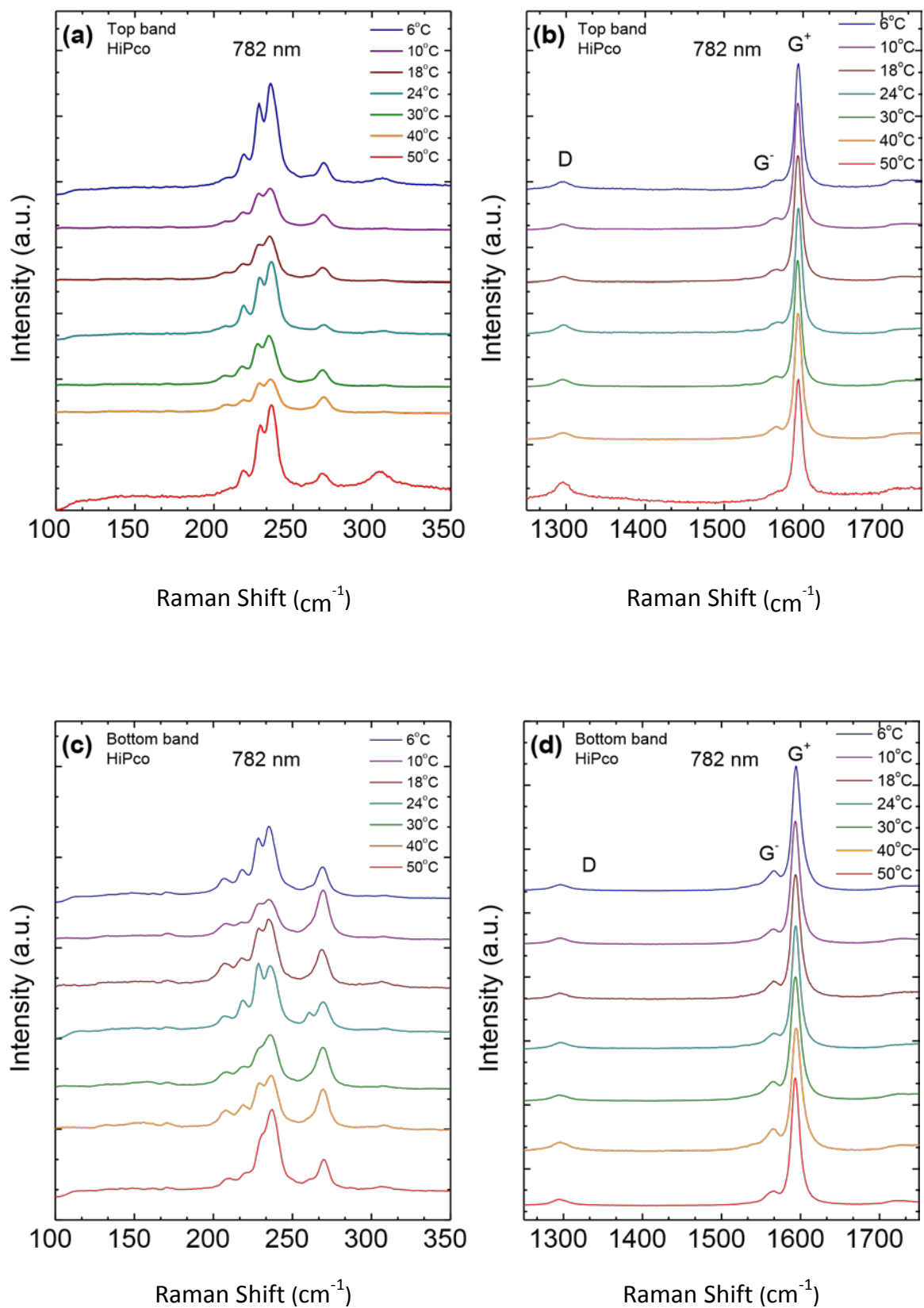

Fig. S3.3: Raman spectra of HiPco-SWNTs separated at different temperatures; excitation wavelength 782 nm. (a) RBM and (b) D-G regions for top band; (c) RBM and (d) D-G regions for bottom band.

## S4. Raman spectra of CoMoCAT-SWNTs

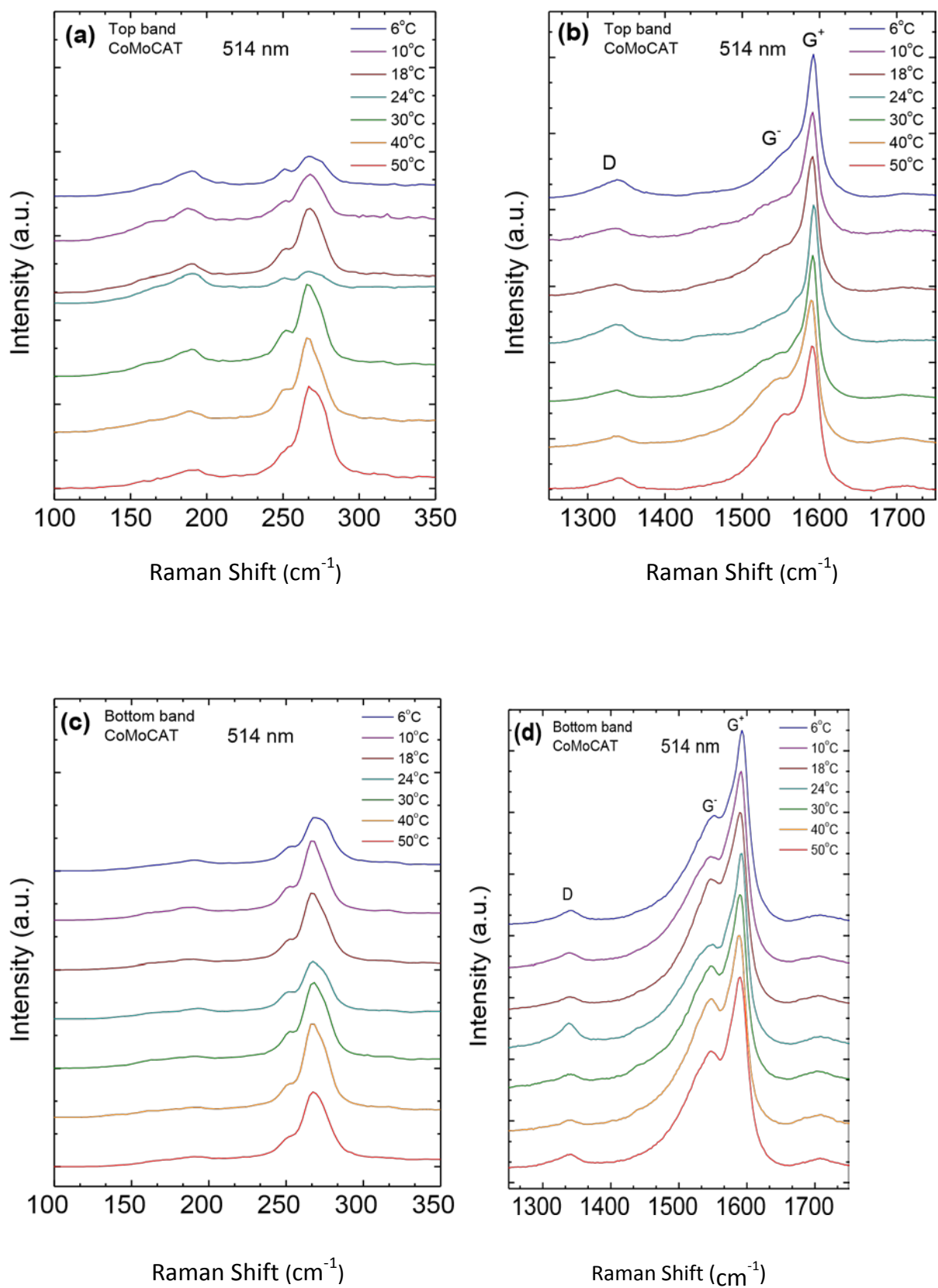

**Fig. S4.1**: Raman spectra of CoMoCAT-SWNTs separated at different temperatures; excitation wavelength 514 nm. (a) RBM and (b) D-G regions for top band; (c) RBM and (d) D-G regions for bottom band.

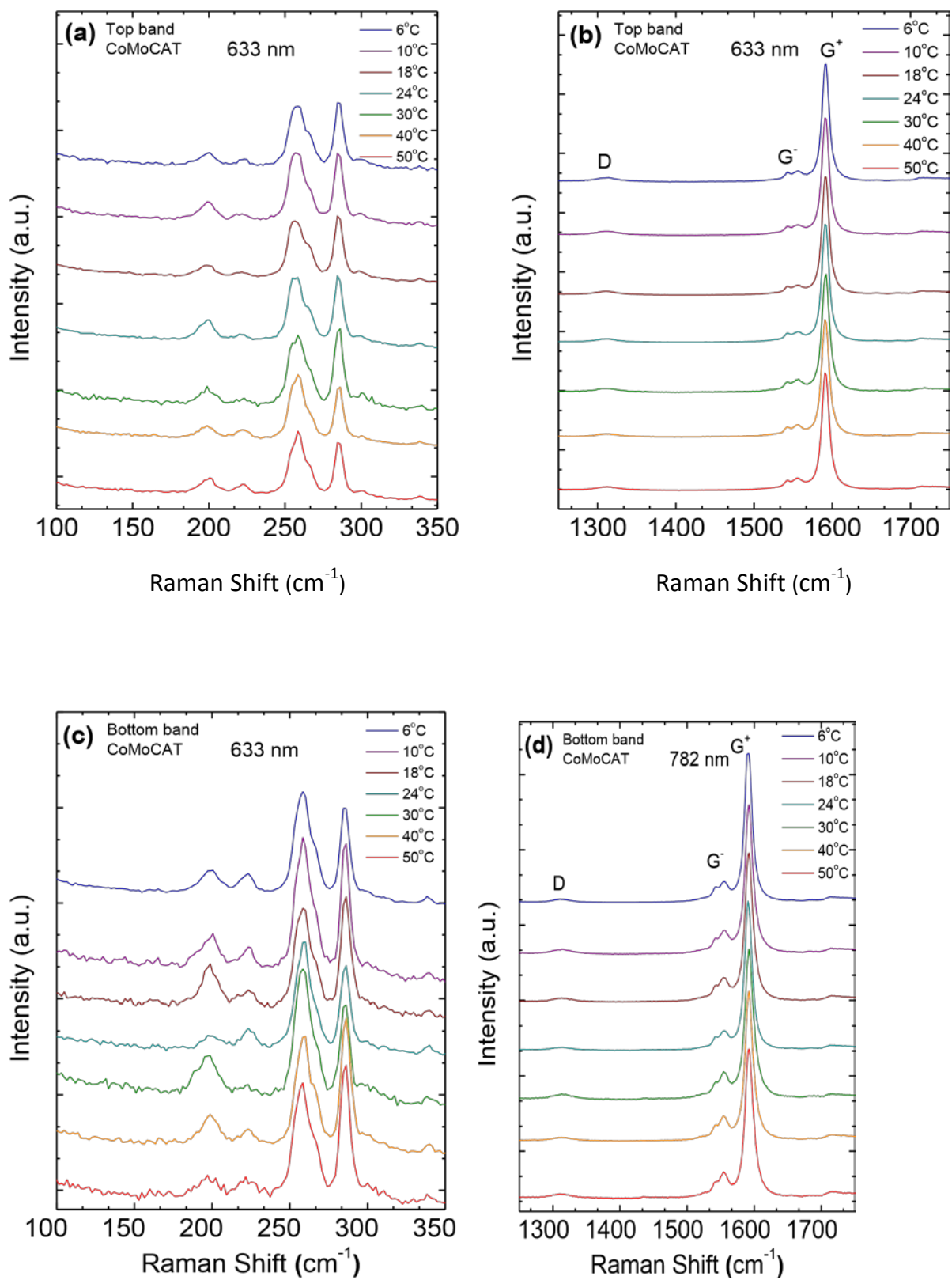

**Fig. S4.2**: Raman spectra of CoMoCAT-SWNTs separated at different temperatures; excitation wavelength 633 nm. (a) RBM and (b) D-G regions for top band; (c) RBM and (d) D-G regions for bottom band.

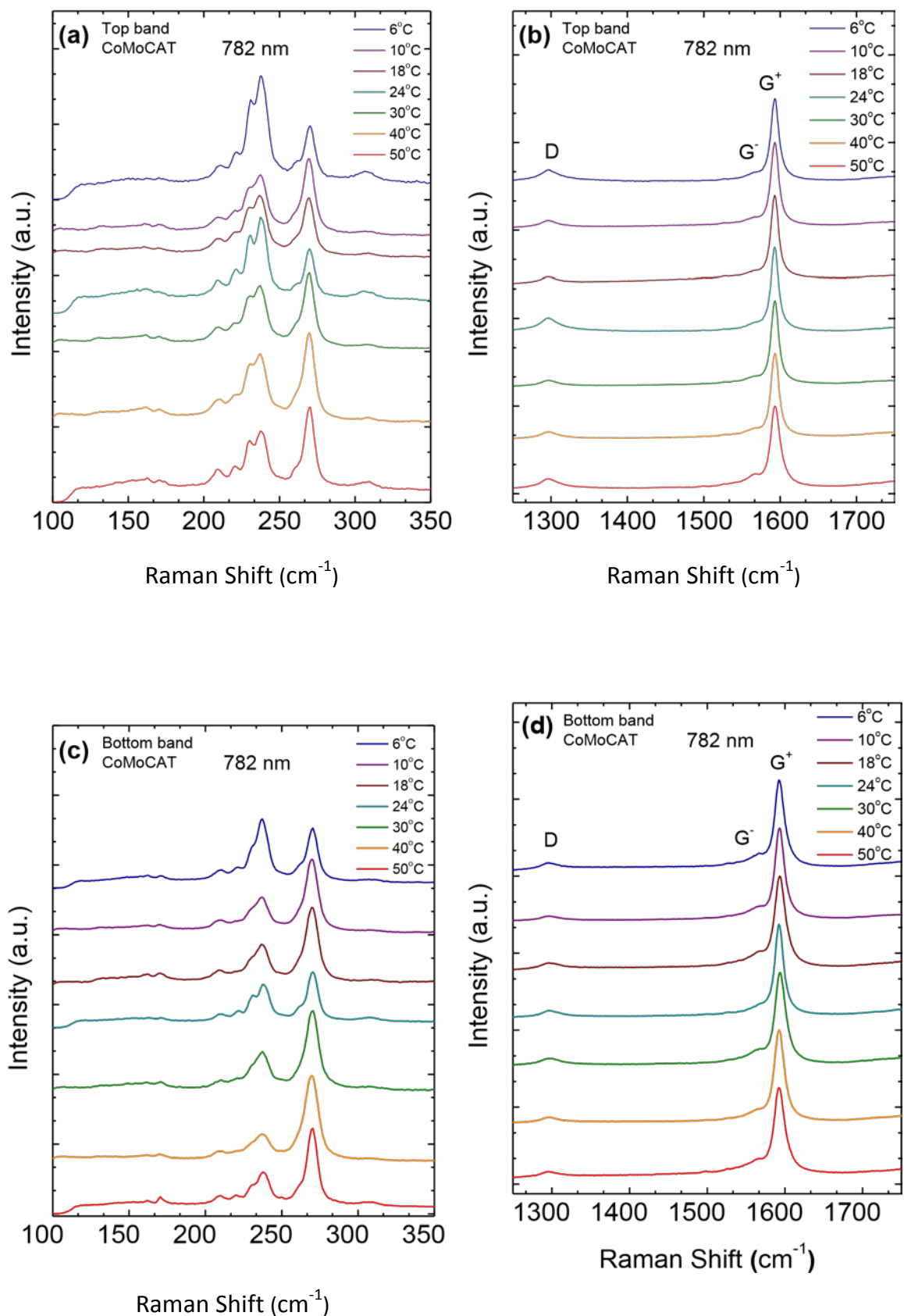

Fig. S4.3: Raman spectra of CoMoCAT-SWNTs separated at different temperatures; excitation wavelength 782 nm. (a) RBM and (b) D-G regions for top band; (c) RBM and (d) D-G regions for bottom band.

## S5. Relative SWNT concentrations in the top and bottom bands after separation

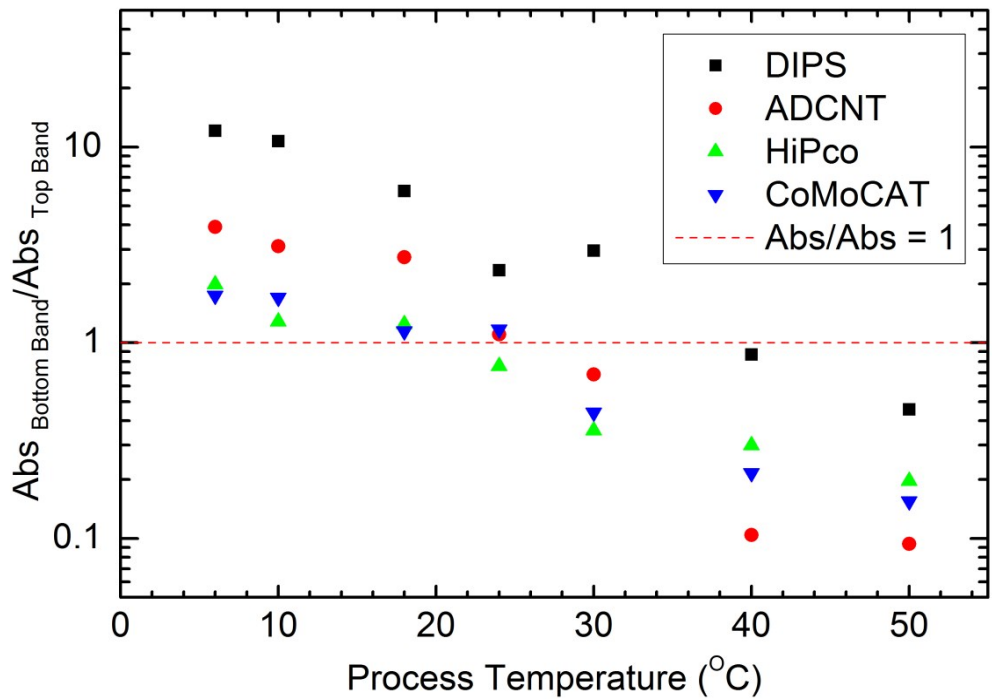

**Fig. S5**: Ratio of optical absorption of the bottom band to the top band as a function of process temperature. The optical absorption is directly proportional to the SWNT concentration in the respective bands.

Fig. S5 reports the variation of the optical absorption with respect to temperature for each sample at wavelengths corresponding to the optical transition boundaries, i.e. between the excitonic transitions. The absorbance at these wavelengths is the background contribution due to the carbon π-plasmon, which is directly proportional to the SWNT concentration.